\documentclass[aps,prd,reprint,showpacs,amsmath,amssymb,superscriptaddress]{revtex4-1}

\usepackage{graphicx}
\usepackage{dcolumn}
\usepackage{bm}
\usepackage{graphics}
\usepackage{graphicx}
\usepackage{slashed}
\usepackage{cancel}
\usepackage{color}

\usepackage{bm}
\usepackage[colorlinks=true,citecolor=cyan,urlcolor=blue,bookmarks=true,bookmarks=true,bookmarksopen=true,bookmarksnumbered=true,bookmarksopenlevel=3]{hyperref}

\definecolor{airforceblue}{rgb}{0.36, 0.54, 0.66}
\definecolor{steelblue}{rgb}{0.27, 0.51, 0.71}
\definecolor{amber}{rgb}{1.0, 0.49, 0.0}

\begin{document}

\title{Production of exotic composite quarks at the LHC}

\author{\textsc{O.~Panella}} 
\affiliation{Istituto Nazionale di Fisica Nucleare, Sezione di Perugia, Via A.~Pascoli, I-06123 Perugia, Italy}
\email[({\bf Corresponding Author})\\ Email: ]{orlando.panella@pg.infn.it}
\author{\textsc{R.~Leonardi}} 
\affiliation{Dipartimento di Fisica e Geologia, Universit\`{a} degli Studi di Perugia, Via A.~Pascoli, I-06123, Perugia, Italy}
\affiliation{Istituto Nazionale di Fisica Nucleare, Sezione di Perugia, Via A.~Pascoli, I-06123 Perugia, Italy}
\author{\textsc{G.~Pancheri}}
\affiliation{Laboratori Nazionali di Frascati, INFN, Frascati Italy}
\author{\textsc{Y.~N.~ Srivastava}}
\affiliation{Dipartimento di Fisica e Geologia, Universit\`{a} degli Studi di Perugia, Via A.~Pascoli, I-06123, Perugia, Italy}
\author{\textsc{M.~Narain}}
\affiliation{Physics Department, Brown University, Providence, USA}
\author{\textsc{U.~Heintz}}
\affiliation{Physics Department, Brown University, Providence, USA}

\date{\today}
\begin{abstract}
We consider the production at the LHC of exotic composite quarks of charge $Q=+(5/3) e$ and $Q=-(4/3) e$. Such states are predicted in composite models of higher isospin multiplets ($I_W=1$ or $I_W=3/2$). Given their exotic charges (such as $5/3$), their decays proceed through the electroweak interactions. We compute decay widths and rates for resonant  production of the exotic quarks at the LHC. Partly motivated by the recent observation of an excess by the CMS collaboration in the $ep_T\mkern-19.5mu\slash\,\,\, jj$
final state signature we focus on $ pp \to U^+ j \to  W^+ + j\, j\, \to \ell^+p_T\mkern-19.5mu\slash\,\,\, jj$
and then perform a fast simulation of the detector reconstruction based on {\scshape{Delphes}}. We then scan  the parameter space of the model ($m_*=\Lambda$) and study the statistical significance of the signal against the relevant standard model background ($Wjj$ followed by leptonic decay of the $W$ gauge boson) providing the luminosity curves as function of $m_*$ for discovery at 3- and 5-$\sigma$ level.  
\end{abstract}

\maketitle
\section{Introduction}
The idea of a further level of compositeness, i.e. that quarks and leptons might not be truly elementary particles~\cite{Dirac:1963aa} but are instead bound states of some as yet unknown entities has been investigated phenomenologically since quite some time back~\cite{Terazawa:1976xx,Terazawa:1979pj,Eichten:1980aa,Eichten:1983hw,Terazawa:1984bd,Cabibbo:1984aa}. One  immediate consequence of this composite scenario is of course that at some high energy scale, the compositeness scale $\Lambda$, excited fermions, quarks and leptons, of mass $m_*\approx \Lambda$ are expected. The interactions of such states with the ordinary quarks and leptons have been modeled on the base of the symmetries of the standard model and are of the magnetic moment type~\cite{Cabibbo:1984aa}.

To the best of our knowledge theoretical and phenomenological studies about the production at colliders of such excited states have concentrated on the multiplets of isospin $I_W=0, 1/2$~\cite{Baur:1990aa,Baur:1987ga}.
The LHC experiments have  produced new interesting results already starting with the early data of Run I providing the  stringent bounds on the mass of excited fermions~\cite{PhysRevLett.105.161801,
Aad:2016aa}, again  restricted to  isospin assignments of $I_W=0,1/2$ for the excited states. 

Two of the present authors studied in ref.~\cite{Pancheri:1984sm} the weak isospin spectroscopy of excited quarks and leptons  showing that the structure of the standard model symmetries allow to consider higher isospin multiplets up to $I_W=1,3/2$. As  a consequence one finds for instance  that the multiplet with $I_W=3/2$ (quartet) contains exotic states such as quarks $U^+$ of charge $+5/3e$  and quarks $D^-$ of charge  $-4/3 e$.

We remind that alternative scenarios beyond the standard model (BSM)  like Little Higgs and Composite Higgs~\cite{Matsedonskyi:2016aa} models predict the existence of vector-like quarks~\cite{Aguila:1983aa,Aguilar-Saavedra:2009aa,Aguilar-Saavedra:2013aa,Aguilar-Saavedra:2013ab} which are color-triplet spin 1/2 fermions whose left- and right-handed components have the same transformation properties under the SU(2) gauge group. Such models also predict the existence of vector-like quarks with exotic charges usually denoted as $T_{5/3}$ and $B_{-4/3}$ or $X$ and $Y$~\cite{Barducci:2015aa,Barducci:2014aa,Cacciapaglia:2015aa}.

Alternative possibilities of high charge high mass quarks, partners of the top quark have also been proposed~\cite{Contino:2008hi} to be tested at the LHC and events of the type $t\bar{t}W^+W^-$ have been discussed.

In \cite{Atre:2011ae} a detailed study of the production at LHC of vector-like quarks with electric charges $q=+(5/3)e$, $q=-(4/3)e$ has been reported showing that at a center of mass energy of $\sqrt{s}=14$ TeV and with 100 fb$^{-1}$ of integrated luminosity a heavy quark mass  of 3.7 TeV  could be reached.

 We may recall recent experimental searches for pair production of vector-like top quark partners of charge q=5/3$e$ ($T_{5/3}$) at LHC both at Run I~\cite{Chatrchyan:2014aa}, where top-quark partners with masses below 800 GeV are excluded at 95\% C.L. assuming that they decay to $tW$, and at Run II~\cite{CMS:2015alb} where a data set of $2.2 \text{fb}^{-1}$ has been used to obtain exclusion limits, at $\sqrt{13}$ TeV, of 960 (940) GeV respectively on the mass of a right-handed (left-handed) $T_{5/3}$.
 
Experimental searches of compositeness, already with earlier data of Run I of the LHC ATLAS \cite{PhysRevLett.105.161801} as well as CMS~\cite{PhysRevLett.105.211801} have put lower limits on the mass of excited quarks, respectively $m_*> 1.2$ TeV and $m_*>1.58$ TeV, from searches in the 2-jet final state. 
 In \cite{CMS-PAS-EXO-16-032} an experimental  search for narrow resonances decaying to di-jets is presented which uses  12.9 fb$^{-1}$ data from Run II of the LHC and excludes excited quarks (with standard isospin and electric charges) with masses below  $m_* \approx 5.4$ TeV extending a previous mass limits of $m_* \approx 5.0$ TeV based on a 2.4 fb$^{-1}$ data samples~\cite{Khachatryan:2016ab}.  Similar searches performed by the ATLAS Collaboration with  3.6 fb$^{-1}$ of proton-proton collisions at $\sqrt{s}=13$ TeV        
 report a heavy quark mass limit of 5.2 TeV~\cite{Aad:2016ab}.  We also quote phenomenological studies  of searches at the LHC in the diphoton~\cite{Bhattacharya:2007aa} and $\gamma$-jet~\cite{Bhattacharya:2009aa} final state signatures, showing that in the simplified scenario ($\Lambda=m_*$) with an integrated luminosity of ${\cal O}(200)\, \text{fb}^{-1} $excited quark masses up to $\approx 5$ TeV can be probed (i.e. observed or excluded) at 3$\sigma$ level.
 
Further details about mass limits and searches through other signatures of the excited quarks (and leptons) can be found in the excellent review~\cite{Golling:2016aa} about  Run I, and earlier Run II, searches at LHC for exotic particles.
  
It is important to realize however that the above strong limits on the masses of the excited quarks do apply within the hypothesis of standard weak isospin assignments ($I_W=0,1/2$). Excited quarks belonging to higher multiplets ($I_W=1,3/2$) be their electric charge exotic or not, do not couple to the gluon field (which has $I_W=0)$, and thus it is not possible to produce them resonantly via quark-gluon scattering~\cite{Pancheri:1984sm} (see also discussion below).
 
 The possibility of excited quarks was suggested in the context of the SSC \cite{Eichten:1983hw} but only within the  hypothesis of isospin singlets and doublets ($I_W=0,1/2$).  They were then rediscussed at length in \cite{Baur:1987ga}. On the other hand, during the early days of the $S{\bar p}pS$ collider \cite{Pancheri:1984sm} the possibility of possible exotic final states within higher isospin multiplets ($I_W=1,3/2$)  were explored in $p {\bar p}$  collisions. 
 The phenomenology at high energy colliders of the exotically charged excited fermions from a composite scenario with extended isospin  multiplets has however  been neglected for a long time after they had been pointed out in \cite{Pancheri:1984sm}. Only recently these exotic fermions have received some attention with respect to the lepton sector. Indeed some of the present authors have studied the production at LHC of the doubly charged leptons of the $I_W=1,3/2$ multiplets~\cite{Biondini:2012ny,Leonardi:2014aa}. We should also mention a recent work where the production of the exotic doubly charged leptons at the linear collider has been considered~\cite{Biondini:2015aa} and another one~\cite{Leonardi:2016aa} where  the authors explore the possibility that the excited neutrino ($\nu^*$) is of the Majorana type and study the corresponding like sign dilepton signature at the LHC. Incidentally the theoretical possibility discussed in \cite{Leonardi:2016aa} has been experimentally searched for by the CMS Collaboration and through an analysis of the 2015 data~\cite{CMS-PAS-EXO-16-026} of Run II a heavy composite Majorana neutrino is excluded up to $m_*\approx4.35$ TeV for a value of the compositeness scale fixed at $\Lambda=5$ TeV. 
 
\begin{figure}[t]
\includegraphics[scale=0.5]{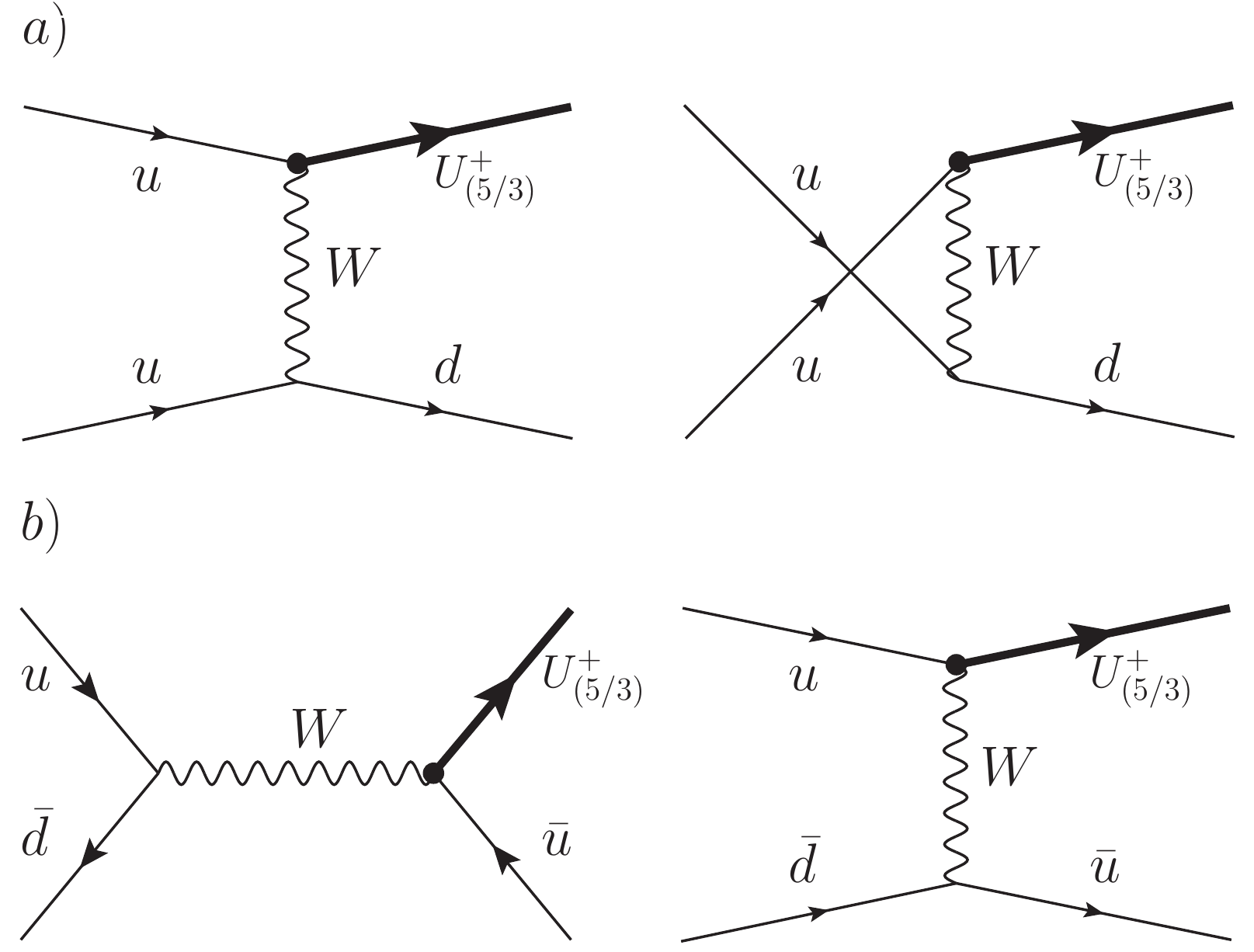}
\caption{\label{fig:ustar} Example processes of $U^+$ resonant production in $pp$ collisions: $a)$ the Feynman diagrams for the process $u u \to U^+ d$ showing explicitly   $W$ exchange in the $t$-channel (left) and $u$-channel (right); $b)$  the Feynman diagrams contributing to the process $ u \bar{d} \to U^+  \bar{u}$ . On the left the $s$-channel annihilation. On the right $t$-channel $W$ exchange. The heavy line is the exotic quark $U^+$, the heavy dot is the magnetic type coupling characteristic of excited quarks (of electroweak strenght in this case). 
}
\end{figure} 

Taking up the model in \cite{Pancheri:1984sm}, here we focus on the phenomenology of quarks with higher charges, belonging to weak isospin quartets, $I_W=3/2$ and $I_W=1$, which contain quarks with charges $5/3$ as well as $-4/3$ (apart from states with standard quark charges $+2/3$ and $-1/3$). The interest of these excited quarks lies in the possibility that they belong to a family whose first generation particles would be coupled to the  $u$-quarks of the standard model (SM).
A peculiar property of the exotic states $U^+$ of charge $q=5/3e$ is that they couple only to the $W$ gauge boson~\cite{Pancheri:1984sm}.  This can be understood by applying the standard rules of addition of quantum angular momentum and by keeping in mind that that the electroweak gauge bosons $(B_\mu, \vec{W}$) have respectively $I_W=0,1$. Then by recalling that the standard model quarks and leptons appear in isospin singlets ($I_W=0$) and doublets ($I_W=1/2$) it is clear that  the excited triplet ($I_W=1$) can only couple to the SM singlet via the gauge field $\vec{W}$. Similarly the ($I_W=3/2$) multiplet can only couple to the SM doublet again via the $\bm{W}$ gauge field. By the same token the gluon field which has ($I_W=0$) cannot couple to a  transition current between the higher isospin multiplets ($I_W=1,3/2$) and the standard model particles. 
 We conclude these considerations by noting that in our composite scenario  the direct coupling of the excited quarks (and leptons) to the SM gauge bosons $\gamma, Z,g$, (e.g. $\gamma, Z, g \to q^* \bar{q}^*$) are expected to be highly suppressed by the presence of form factors. This would be very much similar to what happens in nuclear physics with nucleus-antinucleus pair production which is strongly suppressed at $Q^2 \approx m_A^2$ even if the nucleus has a huge electric charge ($Ze$, with $Z \gg 1$). This implies that the bounds on vector top partners  based on pair production of $T5/3$~\cite{Chatrchyan:2014aa,CMS:2015alb} cannot be applied to the $U^+$ quark of the present model that  are singly produced.

The exotic quarks $U^+$ have only one decay channel $U^+ \to W^+ u$ with ${\cal B} (U^+ \to W^+ u )=1$. 
This implies that they could be resonantly produced via the $2 \to 2 $ process $u u \to U^+ {d}$  and hence decay with unit probability  to $W u$. 

We will discuss therefore the production of the \emph{exotic} excited quarks $U^+, {D}^-$ at the LHC: 
\begin{subequations}
\label{signal}
\begin{align} 
p p \to U^+ j\, ,\\
p p \to D^- j\, ,
\end{align}
\end{subequations}
and finally assuming the leptonic decay of the $W$-gauge boson $W\to \ell \nu_{\ell}$ we concentrate on the $\ell\,\slashed{p}_T\, j\,j $ signature(s):
\begin{subequations}
\label{signature}
\begin{align}
\label{signature1}
 pp \to U^+ j \to  W^+\, j\, j\,\to \ell^+\,\slashed{p}_T\, j\,j\, ,\\
\label{signature2}
pp \to D^- j \to \ell^-\,\slashed{p}_T\, j\,j\,.
\end{align}
\end{subequations}

The production of the state of charge $Q=5/3 e$ has the largest production cross-section for a $pp$ machine such as the LHC due to the availability of two valence $u$ quarks from  the colliding particles. The production of $4/3$ states requires a $\bar{d}$ from the sea and hence its rate is somewhat lower.


We show in Fig.~\ref{fig:ustar} the Feynman diagrams describing some of the  parton sub-process contributing  Eq.~\ref{signature1} which can produce such exotic final state, $U^+(5/3)$. The processes in Eq.~\ref{signature2} will be given by similar diagrams and will  involve at most two valence $d-$quarks in the initial state and thereby the corresponding  production cross section are expected to be somewhat smaller than those of the processes in Eq.~\ref{signature1}.  Here it will be the $dd$ initiated process that dominates ($t$ and $u$-channel exchange of a $W$).

Our phenomenological study of the production of exotic excited quarks in the $\ell\,\slashed{p}_T\, j\,j $ channel is particularly interesting in view of the recent claim of the CMS collaboration of having observed excesses, relative to the standard model (SM) background,  in the data of the Run I at the LHC at $\sqrt{s}=8$ TeV in  the $e e jj $ and $e\,\slashed{p}_T\, j\,j $ channels. 
Indeed the analysis in~\cite{Khachatryan:2014dka} for a search of right-handed gauge boson, $W_R$, based on 19.7 fb$^{-1}$ of integrated luminosity collected at a center of mass energy of 8 TeV reports a 2.8$\sigma$ excess in the ${eejj}$ invariant mass distribution in the interval $1.8\, \text{TeV} < M_{eejj} < 2.2\, \text{TeV}$. 
A CMS search~\cite{CMS-PAS-EXO-12-041b,*Khachatryan:2016aa} for first generation lepto-quarks at a center of mass energy of 8 TeV and 19.6 fb$^{-1}$ of integrated luminosity reported an excess of  2.4$\sigma$ and 2.6$\sigma$ in the $eejj$ and $ep_T\mkern-19.5mu\slash\,\,\, jj$ channels respectively.

The absence of a corresponding excess in the $\mu\slashed{p}_T jj$ channel, as reported in \cite{CMS-PAS-EXO-12-041b,*Khachatryan:2016aa}, will be difficult to explain solely  in terms of heavy exotic quark $U^+$ or $D^-$ resonant production, via the processes in Eqs~(\ref{signature1},\ref{signature2})
,  because the lepton comes from the $W$ gauge boson and thus electrons and muons will have the same yield. However within our composite fermions scenario   
the signature $\ell\slashed{p}_T jj$ could get a contribution also from an  excited neutrino $\nu_\ell^*$ being  produced in association with a lepton $pp \to \ell \nu_\ell^*$ and then decaying as $\nu_\ell^*\to \nu_\ell Z \to \nu_\ell jj$. 
One could therefore qualitatively  explain the fact that the  excess is observed only  in the $e\slashed{p}_T jj$ via the combined production and decay of a heavy composite exotic quark $U^+$ and an excited neutrino by simply assuming that the $\nu_\mu^*$ has a higher mass than  $\nu_e^*$. 

We perform a detailed fast simulation of  signal and SM background via the {\scshape{Delphes}} package~\cite{deFavereau:2013fsa} and obtain luminosity curves, with the statistical error, as function of the parameter ($m_*$) at the 3- and 5-$\sigma$ level. We find that  for different values of the integrated luminosity: (30,300,3000) fb$^{-1}$, commonly used in the study of the LHC  Run II ($\sqrt{s}=13$ TeV) searches, the corresponding mass discovery reach at  the 3-$\sigma$ level is respectively $m_*\approx (2800,3500, 4200)$ GeV for the more favourable case $I_W=3/2$.

Our study shows clearly that a full fledged analysis of the upcoming data from the Run II of LHC at $\sqrt{s}=13 $
TeV has the potential of observing the signature or alternatively excluding larger values of the exotic heavy quark masses ($m_*$) compared to those values already excluded from analyses of Run I~\cite{Aad:2013ab,CMS:2015jga} but applicable only to the  standard excited quarks (with non-exotic charges).

The rest of the paper is organized as follows: In Sec.~\ref{sec:model} we review the theoretical composite model; in Sec.~\ref{sec:productionxsections} we discuss the heavy exotic quark  production cross sections and decay rates; in Sec.~\ref{sec:sig_bg} we discuss the $\ell\slashed{p}_Tjj$ signature and the main associated standard  model background and discuss the kinematic cuts needed to optimize the statistical significance; in Sec.~\ref{sec:fast_simulation} we present the results of the fast simulation obtained through the {\scshape Delphes}~\cite{deFavereau:2013fsa} software and present the 3- and 5-sigma luminosity curves in the parameter space; finally Sec.~\ref{sec:disc_conc} gives the conclusions with outlooks. 

\section{The extended weak-isospin model}
\label{sec:model}
It is well known that in hadronic physics the strong isospin symmetry allowed to discover  baryon and meson resonances well before the observation of quarks and gluons. The properties of the hadronic states could be delineated using the SU(2) and SU(3) symmetries. In analogy with this it may be expected that, for the electroweak sector, the weak isospin spectroscopy could reveal some properties of excited fermions without reference to a particular internal structure.

The standard model fermions have $I_W=0$ and $I_W=1/2$ and the electroweak bosons have $I_W=0$ and $I_W=1$, so, combining them, we can consider fermionic excited states with $I_W\leq 3/2$. The multiplets with $I_W=1$ (triplets) and $I_W=3/2$ (quadruplet) of the hadronic sector include the quarks of exotic charges that are studied in this work: 
\[ \text{U} = \left( \begin{array}{l}
U^{+} \\
U \\
D \end{array} \right) ,
\quad \text{D} = \left( \begin{array}{l}
U \\
D \\
D^{-} \end{array} \right)\,, \quad
\Psi = \left( \begin{array}{l}
U^{+} \\
U \\
D \\
D^{-} \end{array} \right) \, ,\] 
with similar multiplets for the antiparticles.
While referring to the original work in \cite{Pancheri:1984sm} for a detailed discussion of all couplings and interactions, we discuss  here only the main features of these higher multiplets.   We refer to \cite{Biondini:2012ny} for further details and here we mention only  that the higher isospin multiplets ($I_W=1,3/2$) contribute solely to the iso-vector current and do not contribute to the hyper-charge current. 
\begin{figure}[t]
\includegraphics[scale=0.625]{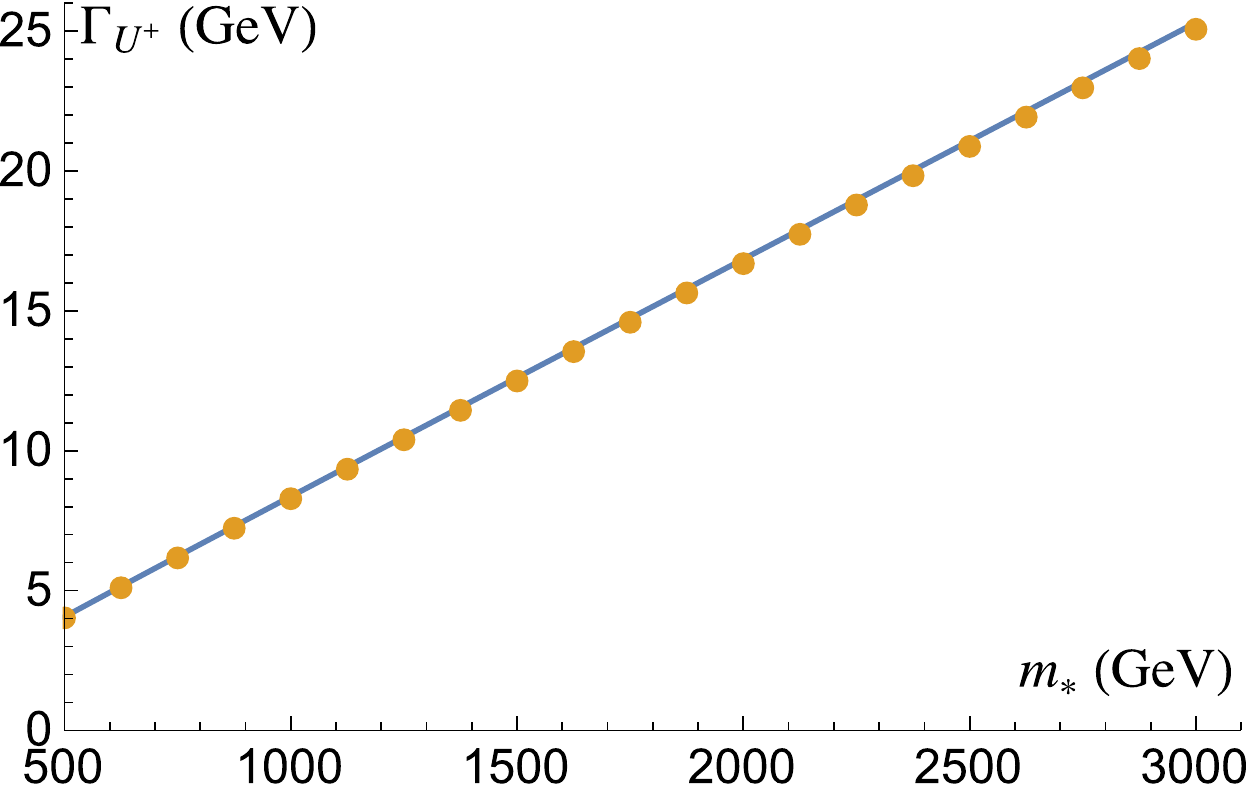}
\caption{\label{fig:width} (Color online) The width of the exotic quark $U^+$ as a function of its mass. The solid (blue) line is the analytical result in Eq.~\protect\ref{width_eq} which is compared with the CalcHEP output, dots (orange) as obtained form the implementation of our model. The agreement is excellent. }
\end{figure}
In order to calculate in detail production and decays of these excited 
fermions, we need to discuss the nature of their couplings to light 
fermions and the gauge fields. Because all the gauge fields carry no
hyper-charge  $Y$, a given excited multiplet couples (through the
gauge field) only to a light multiplet with the same $Y$. Also the
coupling has to  be of the anomalous magnetic moment type, for
current conservation. The  decay modes and reaction cross sections
can be calculated using the following effective lagrangian in terms of
the transition currents: 
\begin{eqnarray}{\cal L}_{\text{int}}^{(I_W=3/2)}&=&  {g{f_{3/2}}\over{\Lambda}} \sum_{M,m,m'}^{\phantom{M}}\,C ( {{3}\over{2}},M |
1,m;{{1}\over{2}},m' )\times \nonumber \\&\phantom{=}& 
  \left({\bar \Psi}_M \sigma_{\mu\nu}
q_{Lm'}  \right)  \partial ^\nu (W^m)^\mu+ h.c.\\
{\cal L}_{\text{int}}^{(I_W=1)}&=& {g{f_{1}}\over{\Lambda}} \sum_{m=0,\pm1}\left[ \left(\bar{\text{U}}_m \sigma_{\mu\nu}
u_R  \right)\right.  \, +\nonumber \\ &\phantom{=}&\phantom{,} 
\left. \left(\bar{\text{D}}_m \sigma_{\mu\nu} d_R\right)\right]
 \partial ^\nu (W^m)^\mu+ h.c.
\end{eqnarray}
In the above equation $g$ is the SU(2) coupling,  $f_{1}$ and $f_{3/2}$  are unknown dimensionless couplings expected to be of order one. We  will assume them exactly equal to 1 throughout the paper.  The mass of the excited fermions $m_*$ will be assumed to coincide with the compositeness scale $\Lambda$ ($m_*=\Lambda$)  and the  $C$'sare Clebsch-Gordon coefficients.

In particular we see that the particles of these higher multiplets with exotic charges interact with the standard model fermions only via the physical $W$ gauge field. For the exotic quark $U^+$ of charge $q=+(5/3)e$ belonging to the  $I_{W}=1$ triplet and the one of the $I_{W}=3/2$ quadruplet the relevant interaction lagrangians  are respectively:
\begin{subequations}
\label{LagGI}
\begin{align}
\label{lagG32}
\mathcal{L}_{\text{int}\,U^+}^{(I_W=3/2)}=
 \frac{g\,{f_{3/2}}}{\Lambda}\left( \bar{U}^+\sigma_{\mu \nu} \,  \, P_L \, u  \right) \partial^{\nu} \, W^{\mu} + h.c.\\
 \label{lagG1}
\mathcal{L}_{\text{int}\,U^+}^{(I_W=1)}=\frac{g\, f_{1}}{\Lambda}\left( \bar{U}^+ \, \sigma_{\mu \nu} \,  \, P_R \, u \right)\partial^{\nu} \, W^{\mu}  + h. c.
 \end{align}
\end{subequations}
 where: $P_{L}=(1-\gamma^{5})/2$ and $P_{R}=(1+\gamma^{5})/2$ are the chiral projectors and $\sigma^{\mu \nu}=i\left[\gamma^{\mu},\gamma^{\nu} \right] /2$; as usual $g$ is the SU(2) coupling constant $g=e/\sin\theta_W$; the field $U^+$ stands for the exotic quark field both for the case $I_{W}=1$ and $I_{W}=3/2$; $u$ is the  u-quark field.  The effective Lagrangian in (\ref{LagGI}) is a dimension five operator and hence one inverse power of the new physics scale (the compositeness scale) $\Lambda$ appears. In the following phenomenology we will consider the simplified model $\Lambda=m_*$.

With the above interaction Lagrangian we can easily compute the total decay width of the exotic state $U^+$ of 
charge $q=(5/3) e$.
Indeed as it only interacts via the $W$ gauge boson its only decay channel is $U^+ \to W^+ u$, and:
\begin{eqnarray}
\Gamma_{U^+} &=& \Gamma(U^+ \to W^+ u)\cr &=& \alpha_{QED}\, \frac{f_{3/2}^2}{\sin^2\theta_W}\,\frac{m_*}{8}  \, \left (2+\frac{M_W^2}{m_*^2}\right)\,\left(1-\frac{M_W^2}{m_*^2}\right)^2 
\label{width_eq}
\end{eqnarray}  
hence we see that for $m_*\gg M_W$, and assuming $f_{3/2}\sim 1 $: $\Gamma_{U^+}/m_* \approx {\cal O}(\alpha_{QED})$.
This behavior is shown explicitly in Fig.~\ref{fig:width} where the width $\Gamma_{U^+}$ is plotted versus the mass $m_*$. 

We have implemented the interactions of the exotic quarks discussed in section II within the  CalcHEP software~\cite{Pukhov:2004ca}. This has been done with the help of FeynRules~\cite{Christensen:2008py}, a Mathematica package that from a given model lagrangian produces  as output the Feynman rules in a format that can be read by various software tools such as CalcHEP and Madgraph.
we note that the standard model background to the process described in Eq.~\ref{signal} has been discussed in \cite{Belyaev:1998dn,Belyaev:1995yu,Eriksson:2006yt,Eriksson:2007re}.

In Fig.~\ref{fig:width} we give a first comparison of the CalcHEP output within our newly implemented model versus an analytical computation of the width of the exotic massive quark $U^+$. The agreement is excellent. 

\section{Production cross sections
\label{sec:productionxsections}}
The exotic quark $U^+$ interacts with the ordinary quarks  through a typical magnetic type interaction \emph{only via the $W$ gauge boson}  and in $pp$ collisions it can be be produced via the first generation  sub-processes: 
(a) $uu \to U^+\,\, d $ ($t$ and $u$-channel $W$ exchange); 
(b) $u \bar{d} \to U^+\,\, \bar{u} $ ($s$ and $t$-channel $W$ exchange) as depicted in Fig.~\ref{fig:ustar}.
Within the first generation we have the parton sub-processes:
\begin{subequations}
\label{Usub}
\begin{align}
 u \, u \to \ U^+ \, d \to\     W^+u \, d \\
 u \, \bar{d} \to \ U^+ \, \bar{u} \to\     W^+u \, \bar{u} 
\end{align}
\end{subequations} 
which may be observed in either a final state with 4 jets or 2 jets and $W^+$ decaying electroweakly. 
Together with  such a high charge member of the mutiplet,  the lower charge exotic quark member of the multiplet  would also be produced. An exotic excited fermion of charge $Q=-(4/3)e$ may be produced through
\begin{subequations}
\label{Dsub}
\begin{align}
d\, d \to\ D^-\, u \to\ W^-  \, d \, u\\
d \, \bar{u} \to\ D^-\, \bar{d}\to\ W^-  \, { d} \,\bar{d}
\end{align}
\end{subequations}

Similar diagrams to those depicted in Fig.~\ref{fig:ustar} will describe  the production of the exotic state $D^-$. We now discuss the production cross sections of the $2\to2$ processes discussed above. 
\begin{widetext}
\subsection{Partonic cross section}
Following notation and conventions of ref.~\cite{Pancheri:1984sm,Biondini:2015aa} we give here the basic cross-sections of the partonic sub-processes. The case of the weak isospin $I_W=1$ is characterized by the absence of interference between $\hat t-$ and $\hat u-$channel (or $\hat s-$ and $\hat t-$channel).  The partonic cross-section for the processes $u u \to U^+ d$ and $u \bar{d}\to U^+ \bar{u}$ are  given by:
\begin{eqnarray}
\left(\frac{d\hat\sigma}{d\hat{t}}\right)_{uu\rightarrow U^+d}
&=&\frac{1}{4\hat s^2m_*^2}\frac{g^4f_1^2}{16\pi}\left\{\frac{\hat{t} \left[m_*^2(\hat{t}-m_*^2)+2\hat s \hat u+m_*^2(\hat s-\hat u)\right]}{(\hat{t}-M_W^2)^2}+
\frac{\hat u \left[m_*^2(\hat u-m_*^2)+2\hat s\hat{t}+m_*^2(\hat s-\hat{t})\right]}{(\hat u-M_W^2)^2}\right\}\\
\left(\frac{d\hat\sigma}{d\hat{t}}\right)_{u\bar{d}\rightarrow
U^+\bar{u}}&=&\frac{1}{4\hat s^2m_*^2}\frac{g^4f_1^2}{16\pi}\left\{\frac{\hat s \left[m_*^2(\hat s-m_*^2)+2\hat{t}\hat u+m_*^2(\hat{t}-\hat u)\right]}{(\hat s-M_W^2)^2}
+\frac{\hat{t} \left[m_*^2(\hat{t}-m_*^2)+2\hat s\hat u+m_*^2(\hat s-\hat u)\right]}{(\hat{t}-M_W^2)^2}\right\}
\end{eqnarray}
The case of the weak isospin $I_W=3/2$ is characterized instead by  nonzero  interferences between $\hat t-$ and $\hat u-$channel (or $\hat s-$ and $\hat t-$channel) which had been neglected in ref.~\cite{Pancheri:1984sm}.  The partonic cross-sections for the processes $u u \to U^+ d$ and $u \bar{d}\to U^+ \bar{u}$ are  given by:
\begin{eqnarray}
\left(\frac{d\hat\sigma}{d\hat{t}}\right)_{uu\rightarrow
U^+d}=\frac{1}{4\hat s^2m_*^2}\frac{g^4f_{3/2}^2}{16\pi}
\left\{
\frac{\hat{t}\left[m_*^2(\hat{t}-m_*^2)+2\hat s\hat u-m_*^2(\hat s-\hat u)\right]}{(\hat{t}-M_W^2)^2}
+\frac{\hat u \left[m_*^2(\hat u-m_*^2)+2\hat s\hat{t}-m_*^2(\hat s-\hat{t})\right]}{(\hat u-M_W^2)^2}\right.\nonumber\\
+\left.\frac{1}{(\hat u-M_W^2)}\frac{1}{(\hat{t}-M_W^2)}\left(\hat s\hat{t}\hat u+\frac{3}{8}\hat u\hat{t}m_*^2\right)\right\}\\
\left(\frac{d\hat\sigma}{d\hat{t}}\right)_{u\bar{d}\rightarrow
U^+\bar{u}}=\frac{1}{4\hat s^2m_*^2}\frac{g^4f_{3/2}^2}{16\pi}\left\{\frac{\hat s \left[m_*^2(\hat s-m_*^2)+2\hat{t}\hat u-m_*^2(\hat{t}-\hat u)\right]}{(\hat s-M_W^2)^2}
+\frac{\hat{t} \left[m_*^2(\hat{t}-m_*^2)+2\hat s\hat u-m_*^2(\hat s-\hat u)\right]}{(\hat{t}-M_W^2)^2}\right. \nonumber\\
+\left.\frac{1}{(\hat s-M_W^2)}\frac{1}{(\hat{t}-M_W^2)}\left(\hat s\hat{t}\hat u+\frac{3}{8}\hat s\hat{t}m_*^2\right)\right\}
\end{eqnarray}
\end{widetext}
The above formulas have also been checked against the results reported in \cite{Biondini:2015aa} by means of using the crossing symmetry.
\begin{figure}[t]
\includegraphics[scale=0.9]{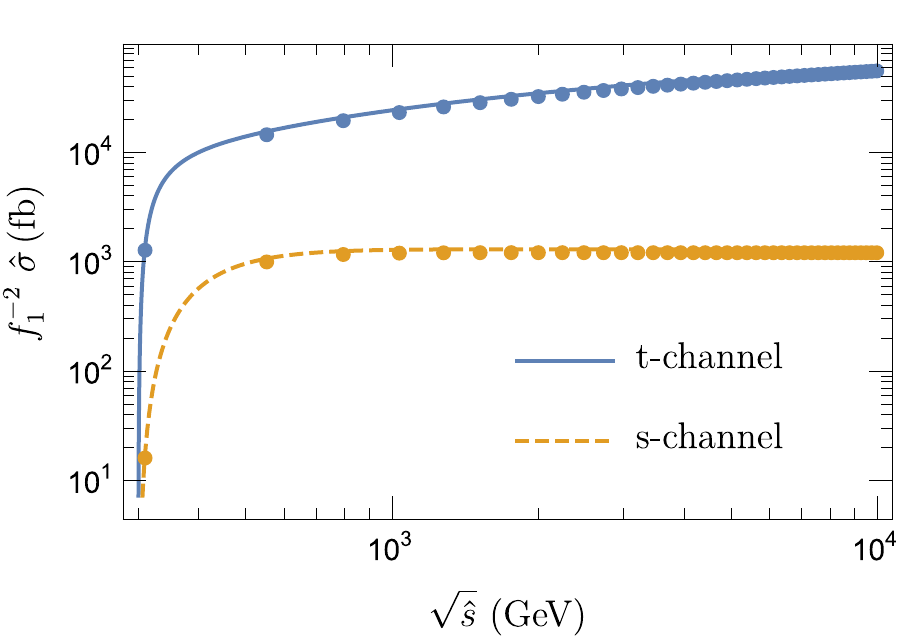}
\caption{\label{fig:fig1} (Color online) For illustrative purposes we give an example of the parton-parton cross-section as indicated in Eq.~\ref{totalpartonic} (no parton ditribution functions). We show $\hat{\sigma} (u\bar{d} \to U^+ \bar{u})$ for $m_*=300$ GeV for the case of the weak isospin $I_W=1$ (in this case there is  no interference between the $t-$ and $s-$channel) and with a choice of the coupling $f_{1q}=1$. The solid line (blue) is the dominant $t$-channel and the dashed line (orange) is the $s$-channel. The dots are the corresponding values obtained by running the same process in CalcHEP within the  model implemented with the help of the FeynRules package. The agreement is excellent, within a few percent.}
\end{figure}
The total integrated cross-section corresponding to the above differential cross section is given for the process $u\bar{d} \to U^+ \bar{u}$ (which receives contributions both from the $s$ and $t$-channels in FIG.~\ref{fig:fig1}. One can see that at high energies the integrated cross section rises logarithmically due to the effect of the $t$-channel $W$ propagator. Also the standard ($1/\hat{s}$) behavior of the cross section is not found because of the magnetic type coupling.
The asymptotic form of the integrated partonic cross sections ($t$-channel) is :
\begin{eqnarray}
\label{totalpartonic}
\hat{\sigma}&=& \int_{-\hat{s}+m_*^2}^{0} d\hat{t} \, \frac{d\hat{\sigma}}{d\hat{t}}\\
&&\quad { \Longrightarrow  \atop \hat{s} \gg M_W^2, m_*^2} \quad  \sim\frac{\pi\alpha_{QED}^2f_{1}^2}{\sin^2\theta_W}\frac{1}{m_*^2} \log(\frac{s}{M_W^2})\nonumber
\end{eqnarray}
as can also be seen from FIG.~\ref{fig:fig1}.
\subsection{Production rates at the LHC}
\begin{figure*}[ht]
\includegraphics[scale=1.4]{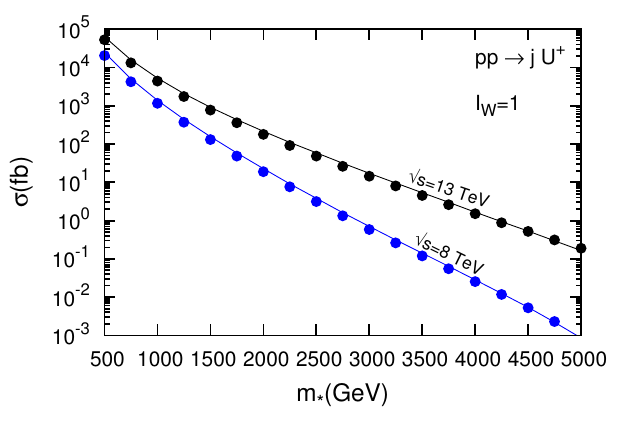}
\includegraphics[scale=1.4]{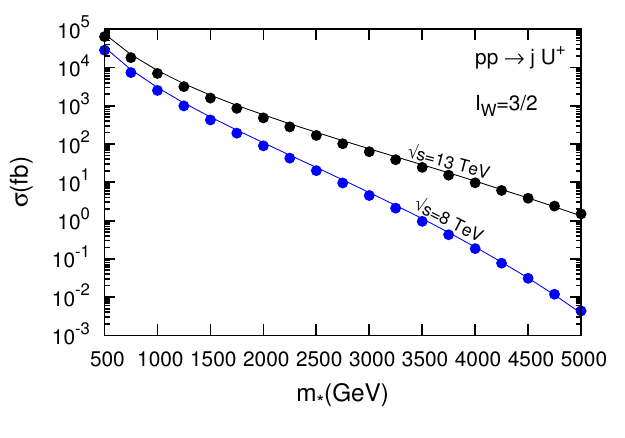}
\caption{\label{fig:xseccompcalchep}(Color online) The total production rates of $p p \to j U^+$ at the LHC for two different values of the energy of the center of mass $\sqrt{s}=8, 13$ TeV and for the two choices of the weak isospin of the exotic states, $I_W =1$ on the left panel and $I_W=3/2$ on the right panel. The solid lines refer to the output of our code based on Eq.~\eqref{prodxsec}. The agreement with of the output obtained with the model implemented in CalcHEP is within a few percent.  Here the factorization and renormalization scale is fixed at $Q=m_*$. The parametrization of the parton distribution function is NNPDF3.0~\protect\cite{Ball:2015aa}.}
\end{figure*} 
\begin{figure*}[ht]
\includegraphics[scale=1.3]{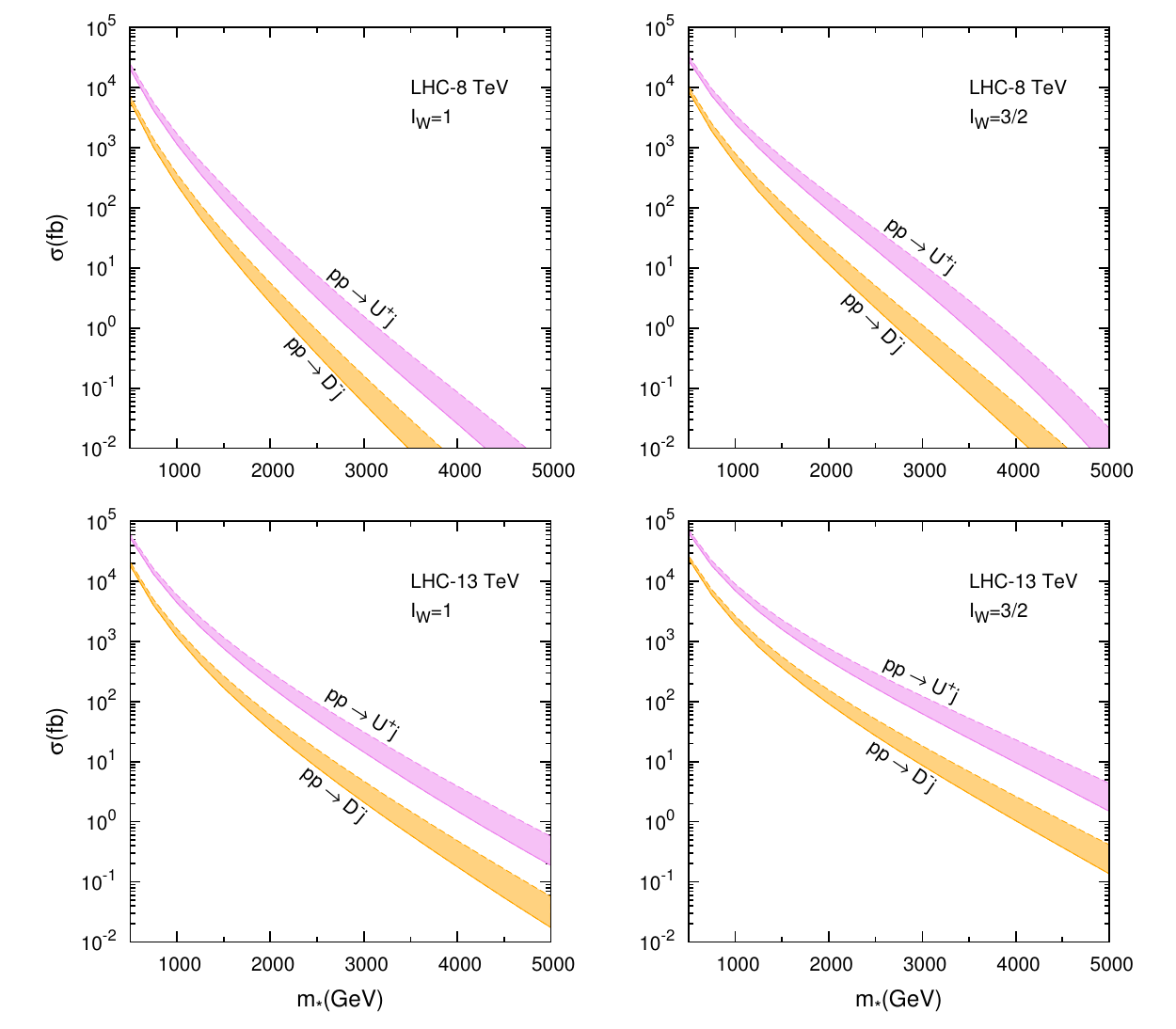}
\caption{\label{fig:xsecprodUD} (Color online) The total integrated cross-sections at the LHC energies of $\sqrt{s}=8, 14$ TeV for the production of the exotic quarks $U^+$, of charge $Q=+(5/3) e$ and $D^{-}$ of charge $Q=-(4/3)e$.
 We have used the NNPDF3.0~\cite{Ball:2015aa} parton distribution functions. The uncertainty bands (magenta and orange) correspond to running the  factorization and renormalization scale from $Q=M_W$ (solid line) up to $Q=m_*$ (dashed line).
All contributing sub-processes within the first two generations (18) have been summed up.}
\end{figure*}

We now present here the production cross sections for the exotic quark $U^+$ in $pp$ collisions expected at the CERN LHC collider according to Feynman's parton model. 
The QCD factorization theorem, allows to obtain the hadronic cross section in terms of convolution of the partonic cross sections $\hat \sigma (\tau s,m_\ast)$,  evaluated at the partons center of mass energy $\sqrt{\hat s}=\sqrt{\tau s}$, and the universal parton distribution functions $f_a$ which depend on the parton longitudinal momentum fractions, $x$, and on the  factorization scale $\hat{Q}$:
\begin{equation}
\label{prodxsec}
\sigma=\sum_{a,b} \,\int_{\frac{m_*^2}{s}}^1\int_\tau^1 \,d\tau \, 
  \frac{dx}{x}\,f_a(x,\hat{Q})\, f_b(\frac{\tau}{x},\hat{Q})\,
\hat{\sigma}(\tau s, m_*)\, .
\end{equation}
In Fig.~\ref{fig:xseccompcalchep} we show a comparison of the production cross sections of $pp\to U^+j$ at $\sqrt{s}=8,13$ TeV between those obtained with an analytical/numerical computation based on Eq.~\ref{prodxsec} (solid line) and those obtained from a CalcHEP numerical simulation based on the implemented model (full dots). The left panel of Fig.~\ref{fig:xseccompcalchep} is for the $I_W=1$ case while the right panel is for $I_W=1$. 

The integrated hadronic cross sections are further shown in Fig.~\ref{fig:xsecprodUD} where we present the results for two different values of the LHC energy, namely $\sqrt{s}=8,13$ TeV. In the top panel
we show for $\sqrt{s}=8$ TeV the total integrated cross section for the production of $U^+$(5/3) and ${D^-}(4/3)$ for $I_W=1$ (left) and $I_W=3/2$ (right).   As expected one finds that the production of $U^+(5/3)$ is  larger. This is almost entirely due to the fact that producing $U^+$ involves the subprocess $u u \to U^+ d$ i.e. with two valence $u$-quarks in the initial state. Similar considerations apply to the results at higher energies (bottom panels). 
For the production of $U^+$ we have, within the first two generations, the following contributing sub-processes:
(a) $u u \to U^+ d$;
(b) $ u \bar{d} \to  U^+ \bar{u}$;
(c) $u c \to U^+ s$;
(d) $ u \bar{s} \to  U^+ \bar{c}$;

\section{Signal and SM Background\label{sec:sig_bg}}

The relevant standard model background to our signature is given by electroweak $Wjj$ production followed by the leptonic decay of the $W$ gauge boson, $W \to \ell \nu_\ell$:
\begin{equation}
pp \to W jj  \to \ell \slashed{p}_T jj
\end{equation} 
This SM background is known to be important and has been discussed throughly in the literature. We have simulated it by using the CalcHEP generator.

We would like to address  here the main kinematic differences between the signal and the relevant SM background in order to choose suitable cuts for optimizing the statistical significance.

One first thing to consider is that one of the two jets is from the heavy quark decay that makes it very energetic with a Jacobian peak in the transverse momentum spectrum near
\begin{equation}
p_T \approx ({m_*}/{2})(1-{M_W^2}/{m_*^2})
\end{equation}

Using the $p_T$ of the jets as a discriminant gives very good accuracy in identifying the jet coming from the decay of the heavy quark correctly, especially for high masses. Hence we identify the hardest jet (j1) in the event as the one from heavy quark decay.

\begin{figure*}[htb]
\includegraphics[scale=1.2]{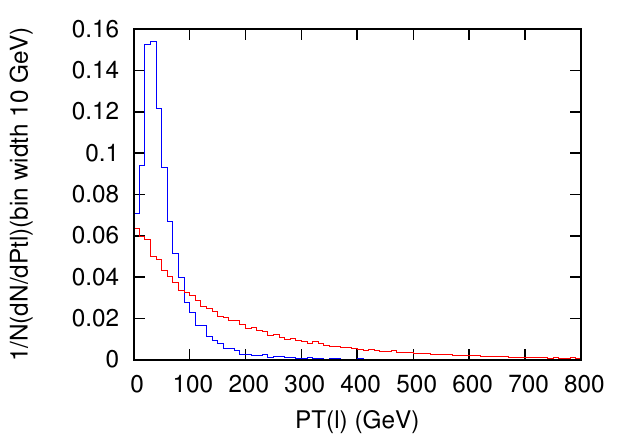}\includegraphics[scale=1.2]{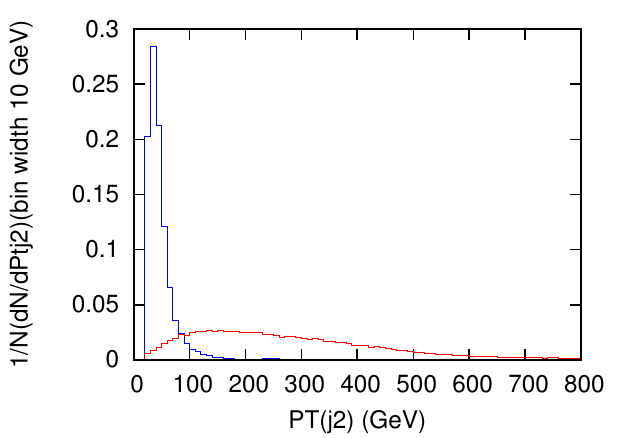}\\\includegraphics[scale=1.2]{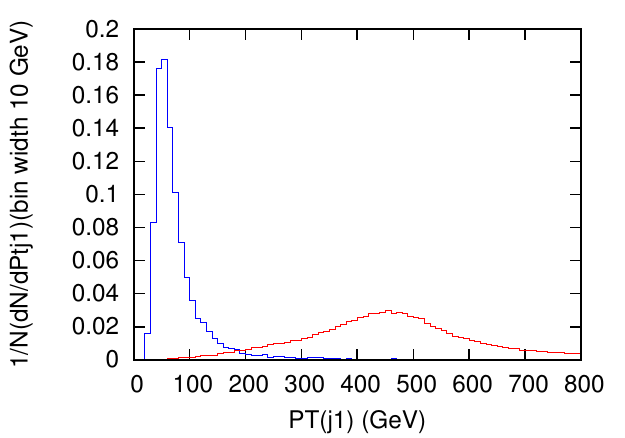}\includegraphics[scale=1.2]{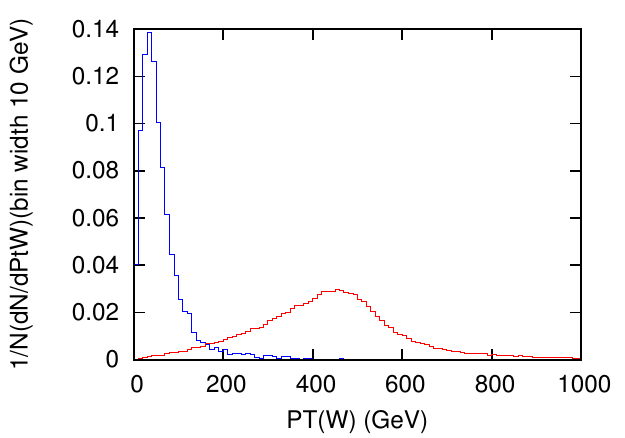}
\caption{\label{fig:ptdistrib} (Color online) Transverse momentum distributions of the signal, dark line (blue), and of the SM background, light line (red): upper left panel, lepton distribution; upper right panel the second-leading jet distribution; lower left panel, leading jet and finally the $W$ gauge boson transverse momentum distribution. These distributions clearly show that the most effective kinematic cut in order to optimize the statistical significance $S$ is one on the  $p_T$ of the leading jet  ($p_{Tj1}\approx O(200)$ GeV that will highly suppress the background while almost will not affect the signal. (d) Transverse $p_T$ of the $W$ gauge boson distribution of our signal $pp \to U^+ \, j \to W ^+ j\, j$ superimposed with the standard model $Wjj$ background.  
}
\end{figure*}

We first define the transverse momentum of the highest $p_T$-jet as $p_{Tj1}$. The main kinematic feature of our signal process is the production of a very heavy excited quark $U^+$ with mass $m_* \approx {\cal O}$ (TeV). At very high masses it will then be a reasonable approximation to assume the exotic heavy particle to be produced nearly at rest. It will decay in a pair of almost back to back high $p_T$ jet and a high $p_T$ W gauge boson. We expect   both the $p_{Tj1}$ and $p_{TW}$ distributions to be peaked at $p_T \approx ({m_*}/{2})(1-{M_W^2}/{m_*^2})$ and to be relatively similar in shape.  These qualitative features are indeed confirmed by our numerical simulation of the signal distributions. Fig.~\ref{fig:ptdistrib} (bottom left and bottom right panels) show the $p_{Tj1}$ and $p_{TW}$ distributions for $m_* =1000$ GeV which are clearly both peaked around $p_T \approx 400 $ GeV in this case.

\begin{figure*}[htb]
\includegraphics[scale=1.3]{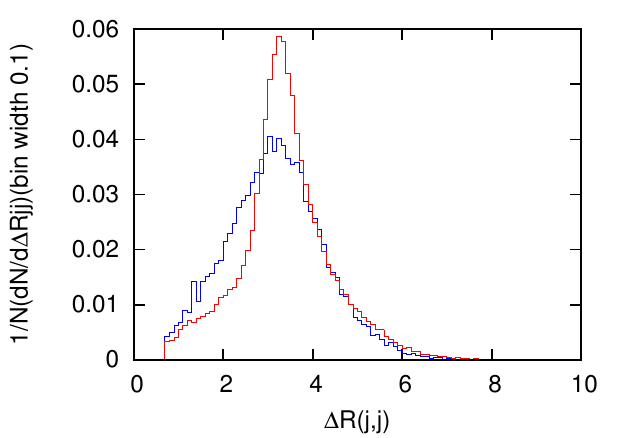}\includegraphics[scale=1.3]{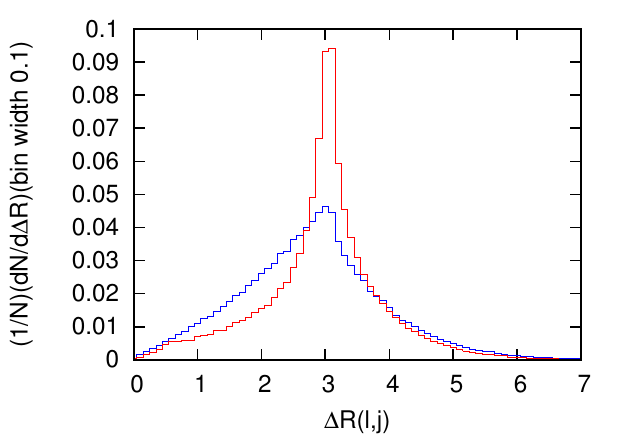}
\caption{\label{fig:DRdistrib}(Color Online) Normalized $\Delta R$ distributions of signal, dark line (blue), and SM background, light line (red), clearly show that the two jets are well separated (plot on the left -$\Delta R_{jj}$-) as well as the lepton and the jets (plot on the right -$\Delta R_{\ell j}$-).}
\end{figure*}
\begin{figure*}[ht!]
\includegraphics[scale=1.25]{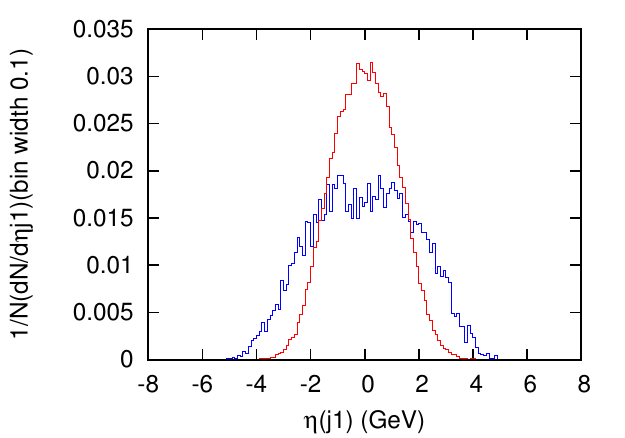}
\includegraphics[scale=1.25]{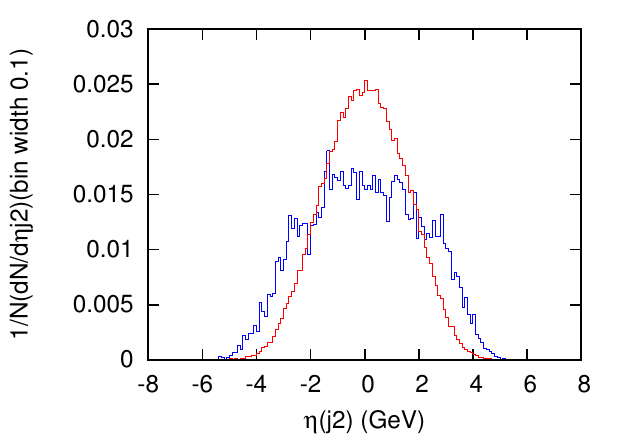}
\includegraphics[scale=1.25]{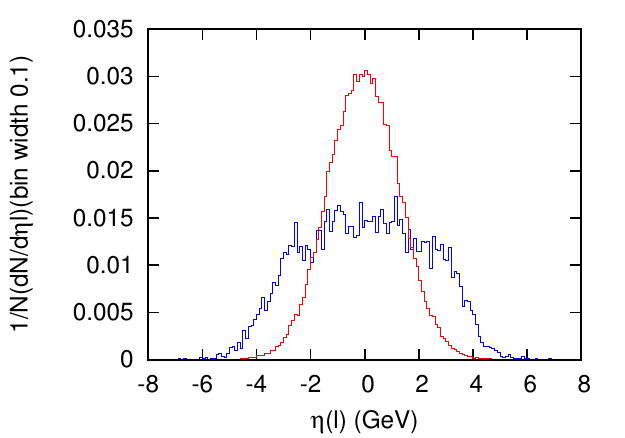}\includegraphics[scale=1.25]{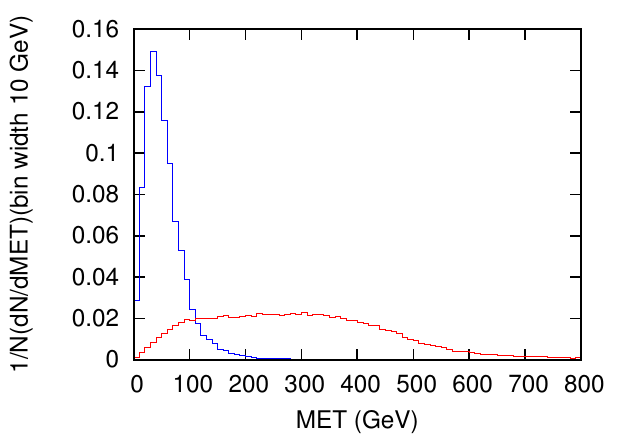}
\caption{\label{fig:angdistrib} (Color Online) Various (normalized) distributions of the signal 
$p p \to e\slashed{p}_T j j$, dark line (blue), in the case  of  an exotic quark state $U^+(5/3) $  mass $m_*= 1000$ GeV and for a compositeness scale $\Lambda=10 $ TeV as well as of the SM background $pp \to W jj  \to \ell \slashed{p}_T jj$, light line (red), at $\sqrt {s} = 13$ TeV. We have used the NNPDF3.0~\cite{Ball:2015aa} parton distribution functions evaluated at the scale $Q=m_*$. N.B. we have considered here all 14 subprocesses for a total of 29 Feynman diagrams within the first generation of quarks. In the top  (left and right) panels and in the lower left panel we show respectively  the pseudo-rapidity distributions of: (a) the highest $p_T$ jet (j1); (b) the second-leading jet (j2); (c) the lepton.   In the bottom right panel we show the missing transverse energy (MET) distribution.}
\end{figure*}
\begin{figure*}
\includegraphics[scale=1.35]{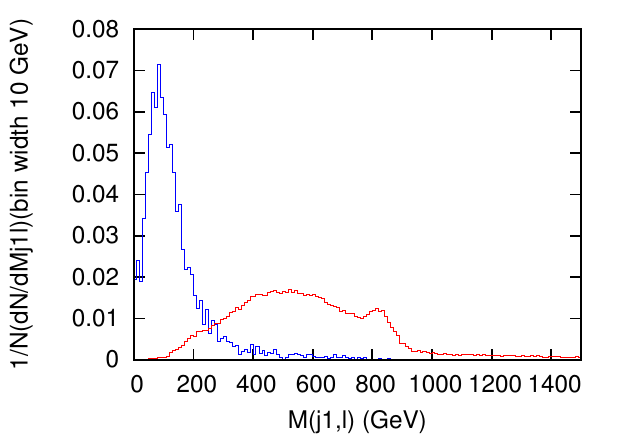}\includegraphics[scale=1.35]{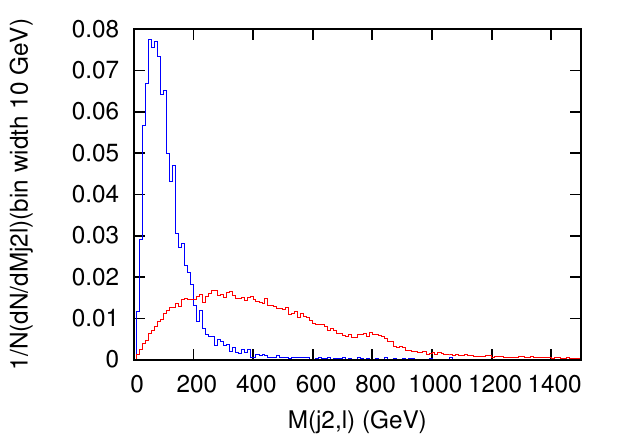}
\includegraphics[scale=1.35]{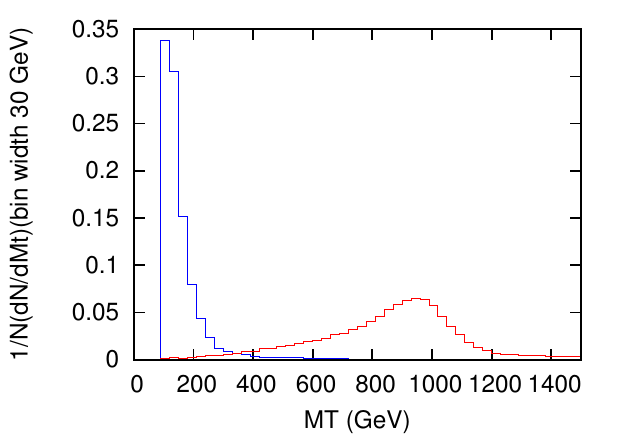}\includegraphics[scale=1.35]{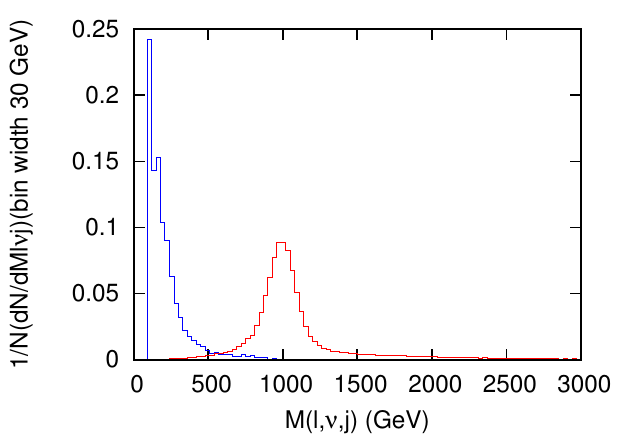}
\caption{\label{fig:inv_mass_distrib} (Color online) Various invariant and transverse mass distributions for a choice of $m_*=\Lambda=1000$ GeV at $\sqrt{s}=13$ TeV for the signal, dark line (blue), and the SM background, light line (red). In the top left and top right panels we show the invariant mass distribution of the lepton with $j_1$ the highest $p_T$ jet (leading) and with $j_2$ the second leading jet. In the bottom panels we show the transverse mass $M_T$ distribution (left) and the reconstructed invariant mass, $M(\ell \nu_\ell j_1)$, of the decay products of the exotic quark $U^+$ (right).}
\end{figure*}

Fig.~\ref{fig:ptdistrib}, Fig.~\ref{fig:DRdistrib}, Fig.~\ref{fig:angdistrib} and  Fig.~\ref{fig:inv_mass_distrib}  show several normalized distributions with respect to both  transverse momentum and angular variables. Fig~\ref{fig:ptdistrib} shows different  transverse momentum distributions:  the transverse momentum  of the  lepton $p_{T\ell}$, the second-leading $p_{T}(j2)$ and that of the leading $p_{Tj1}$ are shown in Fig.~\ref{fig:ptdistrib}(a,b,c) while the  $W$ gauge boson transverse $p_T$ distribution is given in Fig.~\ref{fig:ptdistrib}(d).  

From the point of view of the transverse momentum distributions of the jets (leading and second-leading) in Fig.~\ref{fig:ptdistrib}, signal and background are very well separated, for the given values of the parameters ($m_*=1000$ GeV and $\Lambda=10$ TeV). This suggests that a very efficient way we to reduce drastically the background while keeping most of the signal is a cut  on the  transverse momentum  of the leading jet at $\approx 200$ GeV and, possibly, a cut  in the transverse momentum of the second-leading jet at $\approx 100$ GeV. 

From Fig.~\ref{fig:DRdistrib}   we can see that both for the signal and the background a large fraction of the events have the two jets (or the lepton and the jets) with a large separation in the $(\eta,\phi)$ plane, $\Delta R= \sqrt{(\Delta \eta)^2+(\Delta \phi)^2}$, ($\eta$ is the pseudorapidity and $\phi$ the azimuthal angle in the transverse plane). The corresponding $\Delta R$ distributions are peaked at $(\Delta R)_\text{min} \approx 3$. Therefore, in the reconstruction process,   the two jets can be easily separated as well as the lepton is cleary separated from the jets.

Differently than with the $p_T$ distributions we can see from 
Fig.~\ref{fig:angdistrib}(a,b,c) 
 that the pseudorapidity distributions of the leading jet in $p_T$, $\eta(j_1)$, and the second-leading jet in $p_T$, $\eta(j_2)$, and those of the lepton, $\eta(\ell)$, are quite similar for signal and background.
The missing transverse energy distribution is shown 
Fig~\ref{fig:angdistrib}(d). Here signal and background are quite separated but we have checked that a cut on the missing transverse energy is less effective than one on the transverse momentum of the leading jet, $p_T(j_1)$.  

Let us also  comment on the fact that the
$\ell\nu_\ell jj$ signature from a heavy composite quark state of charge (+4/3)$e$ has the potential to   explain the excess observed in a search for 1$^{st}$ generation lepto-quarks (LQ) by the CMS collaboration~\cite{Khachatryan:2014dka} in the $e\slashed{p}_Tjj$ invariant mass distribution in the interval  $ M_{ej}\approx  600$ GeV. Fig.~\ref{fig:inv_mass_distrib} (a,b) shows, at the reconstructed  level and for a particular point of the parameter space ($\Lambda=m_*=1000$ GeV), that the $ej_1$ (electron and leading-jet) and $ej_2$ (electron and second-leading jet) invariant mass distribution can easily accomodate an excess in the interval where it has been claimed by the CMS collaboration.
Our model would predict the same excess in the muon channel because the leptons arise from the decay of the $W$ gauge boson. Some mechanism would have to be conceived in our model to suppress this excess in the muon channel. As already mentioned in the introduction we  observe that  this signature could also be affected by the  production of an excited neutrino ($N=\nu^*$) in association with the corresponding lepton followed by the decay of the heavy neutrino to a lepton and a gauge boson decaying to two jets, thus obtaining $\ell\slashed{p}_Tjj$. The differences in the electron and muon channels could then be ascribed to a mass hierarchy between the  excited electron and muon  heavy neutrinos.


We point out that within our final state signature ($\ell \nu_{\ell} jj$)  it is always possible to define a \emph{cluster} transverse mass variable ($M_T$) in terms of the reconstructed transverse momentum of the $W$ gauge boson ($\bm{p}_{TW}=\bm{p}_{T\ell}+\bm{p}_{T\nu}$) and the transverse momentum of the leading jet (or highest $p_T$ jet) $p_{Tj_1}$:
\begin{equation}
\label{transverseMASS}
M_T^2=\left( \sqrt{p_{TW}^2 +M_W^2 }+p_{Tj1}\right)^2-\left(\bm{p}_{TW}+\bm{p}_{Tj1}\right)^2
\end{equation} 
The transverse mass distribution is strongly correlated with the heavy exotic quark mass $m_*$. Relevant information about the mass of the heavy exotic quark $U^+$ can be obtained from the transverse mass distribution $M_T$. This is indeed the case as can be seen from Fig.~\ref{fig:inv_mass_distrib}(lower-left) where the transverse mass distribution obtained for the parameter value ($m_*=\Lambda=1000$ GeV) shows a clear peak characterized by a relatively sharp end-point at $M_T\approx m_*$. 
 This is  expected since in the resonant production the heavy exotic quark, $U^+$ is decaying to $\ell \nu_\ell j$ and the jet from $U^+$ is expected to be the leading, while the second-leading jet is the one produced in association with $U^+$, in $pp\to U^+ j$.

Finally we have also reconstructed the invariant mass distribution of the decay products of the heavy excited quark $U^+$: $\ell \nu_\ell j_1$. Indeed it is possible to reconstruct the longitudinal neutrino momentum $p_z(\nu)$ up to a two-fold ambiguity~\cite{Richter:2013aa}. The resulting reconstructed invariant mass $M_{\ell\nu j_1}$ is shown in Fig.~\ref{fig:inv_mass_distrib}(lower-right).   

In order to still reconstruct the invariant mass of the exotic quark to some degree of accuracy, we can follow the method described in \cite{Richter:2013aa} modified to adatpt it to our case. We use the conservation of four-momentum to solve for the longitudinal momentum of the neutrino ($p_L^\nu$). Conservation of four-momentum, $p_W=p_\ell+p_\nu$, gives the following equation:
\begin{equation}\label{mW}
M_W^2=(p_\ell +p_\nu)^2\, .
\end{equation}
The only unknown quantity in Eq.~\ref{mW} is the longitudinal momentum of the neutrino. Expanding the right-hand side of Eq.~\ref{mW} we obtain a second-order equation for $p_L^\nu$:
\begin{equation}
(1-B^2)(p_L^\nu)^2-2A\,B\, p^\nu_T\,p_L^\nu+(p^\nu_T)^2(1-A^2)=0
\end{equation} 
where $p^\nu_T=| \bm{p}_T^{\,\nu}|$ while $p_L^\nu$ and $p_L^\ell$ are the true components (with sign) of the neutrino and lepton momentum along the (longitudinal) $z$-axis and: 
\begin{equation}
A=\frac{M_W^2+2\,{\bm{p}}_T^{\,\ell} \cdot \bm{p}_T^{\,\nu}}{2E_\ell p_T^\nu}\,, \qquad B=\frac{p_L^\ell}{E_\ell}\,.
\end{equation}
It has the solutions:
\begin{equation}
p_L^\nu=\frac{1}{1-B^2}\left[AB\pm\sqrt{A^2+B^2-1} \right]\, p_\nu^T
\end{equation}
Note that the discriminant ($D$) of the second order equation is the quantity in the square root, $D=A^2+B^2-1$. We have three distinct possibilities:
(i) $D>0$, two real solutions;
(ii) $D=0$, one real solution;
(iii) $D<0$, two complex solutions.

If the discriminant is zero there is  only one solution for $p_L^\nu$ which can be used to fully reconstruct the neutrino. If the discriminant is negative, the event is rejected.
If the discriminant is positive, there are two possible $p_L^\nu$ solutions. Using both of them, the two possible neutrino momentum vectors are constructed and, combining them with the lepton momentum, the two $W$ candidate are re-constructed. We select the $p_L^\nu$ solution that gives the more central $W$, i.e. with the smaller pseudo-rapidity. Then we can reconstruct the corresponding  invariant mass $M_{\ell \nu_\ell j1}$. 
Fig.~\ref{fig:inv_mass_distrib} shows the distribution in the invariant mass of the lepton, jet and neutrino. There is a clear peak in correspondence of the exotic quark mass.

\section{Fast detector Simulation and reconstructed objects}
\label{sec:fast_simulation}
\begin{figure*}[bt]
\includegraphics[scale=0.85]{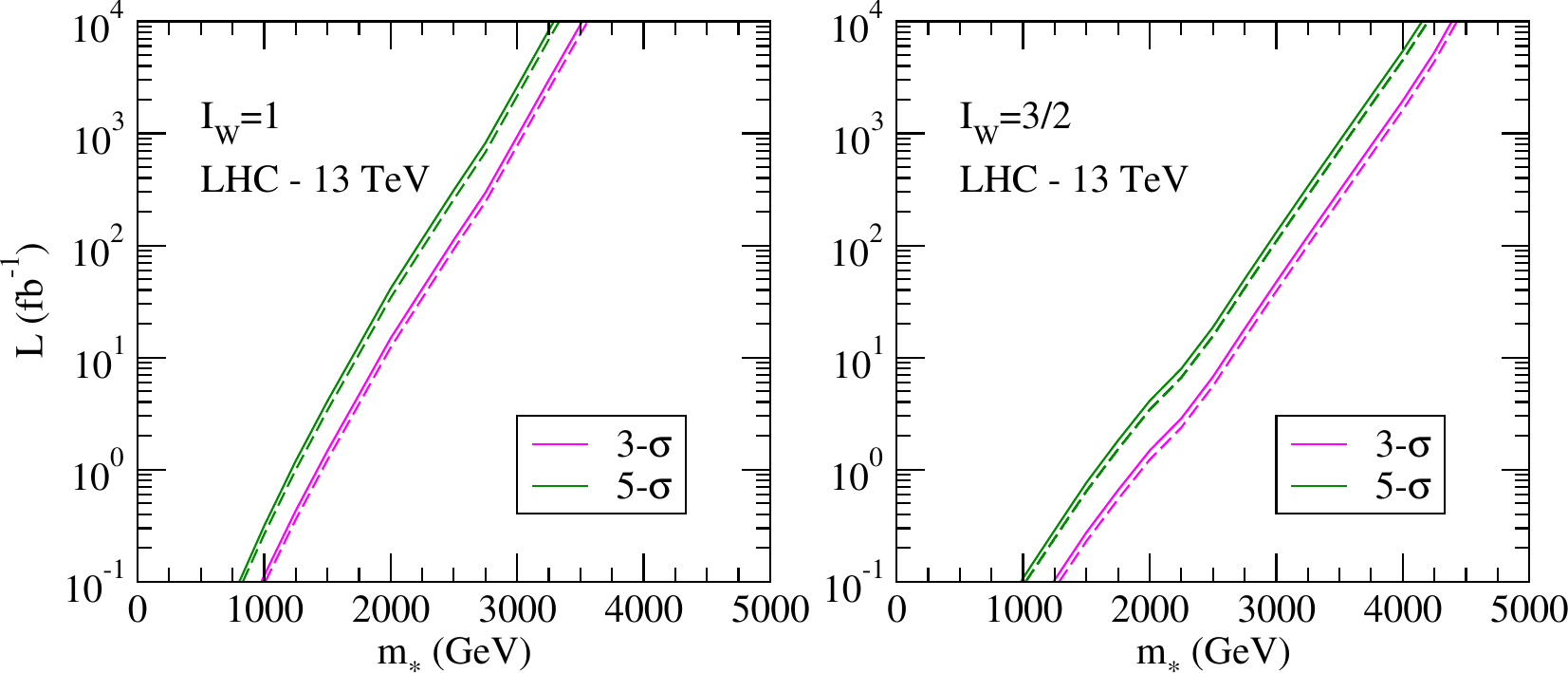}
\caption{\label{fig:lumicurve} (Color online) Luminosity curves at 5-$\sigma$ and 3-$\sigma$ level for $\sqrt{s}=13$ TeV as a function of the excited quark mass $m_*$ after including the fast simulation efficiencies of the detector reconstruction. For  $I_W=1$, and for values of the integrated luminosity equal to $L=(30,300,3000)$ fb$^{-1}$ we find a 3-sigma level   mass reach respectively  up to $m_*\approx (2230,2780,3280)$  GeV. For the same values of the integrated luminosity in the $I_W=3/2$ case we find a  3-sigma level   mass reach respectively  up to $m_*\approx (2930,3540,4140)$ GeV. The solid and dashed lines both in the 3-$\sigma$ and 5-$\sigma$ curves define the band of the luminosity curves within the statistical error.
}
\end{figure*}

In order to take into account the detector effects, such as efficiency and resolution in reconstructing kinematic variables, we interface the LHE output of CalcHEP with the software {\scshape{Delphes}} that simulates the response of a generic detector according to predefined configurations. We use a CMS-like parametrization.
For the signal we consider a scan of the parameter space ($\Lambda=m_*$) within the range  $m_*\in [500,5000]$ GeV with step of $500$ GeV.
\begin{table}
\begin{ruledtabular}
\begin{tabular}{cccc}
cut&$p_T(j1)$&$p_T(j2)$&$M_T$\cr
\hline
t$_1$& $ > 180$  GeV& --&--\cr
t$_2$&$ > 200$ GeV&--&--\cr
t$_3$&$ > 180$ GeV & $ > 100$ GeV&--\cr
t$_4$& $>180$ GeV & -- & $> 400$ GeV\cr
\end{tabular}
\end{ruledtabular}
\caption{\label{tab:cuts}Various cuts which have been studied in order to maximise the statistical significance. It turns out that cut t$_3$ is the most efficient cut.}
\end{table}
We have studied four different choices of kinematical cuts $t_1 \dots t_4$ as described in Table~\ref{tab:cuts}. Although the various choices perform quite similarly, it turns out that the most efficient choice is found to be cut $t_3$: 
\begin{subequations}
\label{cuts}
\begin{align}
\label{Cut1}
p_T(j_{\text{leading}})\geq 180\,\text{GeV},\\
\label{Cut2}
p_T(j_{\text{second-leading}})\geq 100\, \text{GeV}.
\end{align}
\end{subequations}
For each signal point and for the standard model background we generate $10^5$ events in order to have enough statistics to evaluate the reconstruction efficiencies ($\epsilon_s$, $\epsilon_b$) of the detector and of the cuts previously fixed (see Eq.~\ref{Cut1},~\ref{Cut2}).
 We select the events with two jets, one lepton and $\slashed{p}_T$ in the final state. This is justified because the two jets are well separated, as opposed for instance to what happens the signal $pp\to \ell\ell jj$ studied in ref.~\cite{Leonardi:2016aa}, due to a heavy composite Majorana neutrino, where it was found that depending on the heavy neutrino mass ($m_*$) it is possible to have  merging of the two jets in a sizeable fraction of the events. Once we have the number of the selected events we evaluate the reconstruction efficiencies. The efficiencies are shown for the choice of cuts $t_3$ (see Table~\ref{tab:cuts}) in Table~\ref{tab:efficienciesIw3/2}. Then for a given luminosity $L$ it is possible to estimate the expected number of events for the signal ($N_s$) and for the background ($N_b$): 
\begin{equation}N_s=L\sigma_s\epsilon_s\, , \qquad N_b=L\sigma_b\epsilon_b\, , 
\end{equation}
and finally the statistical significance ($S$) is evaluated  as:
\begin{table}[t!]
\begin{tabular}{|c|c|c|c|}
\hline
\multicolumn{4}{|c|}{\textbf{Background}}\\
\hline
 & $\sigma_b$ before cut (fb)& $\sigma_b$ after cut (fb) & ($\epsilon_b$) \\
\cline{2-4}
 & 8200000                 &    14678		     & 0.00179 \\
\hline
\multicolumn{4}{|c|}{\textbf{Signal} ($I_W=1$)}\\
\hline
$m_*$ (GeV)& $\sigma_s$ before cut (fb) & $\sigma_s$ after cut (fb) & ($\epsilon_s$) \\
\hline
500  & 7782 & 5416.74 & 0.69606 \\
1000 & 1277 & 1064.33 & 0.83346 \\
1500 & 344.6 & 298.489 & 0.86619 \\
2000 & 107.7 & 95.1185 & 0.88318 \\
2500 & 39.05 & 34.7037  & 0.8887 \\
3000 & 13.5 & 12.0555 & 0.893 \\
3500 & 4.281 & 3.84352 & 0.89781 \\
4000 & 1.424 & 1.28213  & 0.90037 \\
4500 & 0.4957 & 0.446665 & 0.90108 \\
5000 & 0.1799 & 0.162518 & 0.90338 \\
\hline
\multicolumn{4}{|c|}{\textbf{Signal} ($I_W=3/2$)}\\
\hline
$m_*$ (GeV)& $\sigma_s$ before cut (fb) & $\sigma_s$ after cut (fb) &  ($\epsilon_s$) \\
\hline
500  & 11080 & 5819.11 & 0.52519 \\
1000 & 2240  & 1649.89 & 0.73656 \\
1500 & 806.3 & 646.065 & 0.80126 \\
2000 & 343.2 & 283.964 & 0.8274 \\
2500 & 159.9 & 134.147  & 0.83894 \\
3000 & 60.25 & 51.8626 & 0.86079 \\
3500 & 23.55 & 20.0983 & 0.85343 \\
4000 & 9.347 & 7.57986  & 0.81094 \\
4500 & 3.191 & 2.60797 & 0.81729 \\
5000 & 1.043 & 0.845737 & 0.81087 \\
\hline
\end{tabular}
\caption{\label{tab:efficienciesIw3/2} Efficiencies of the standard model $Wjj$  background and of our signature for the $I_W=3/2$ and $I_W=1$ cases. The estimated efficiencies refer to the choice of kinematic cut t$_3$ described in Tab.~\ref{tab:cuts} or Eqs.~\ref{Cut1},~\ref{Cut2}. }
\end{table}
\begin{equation}
S=\frac{N_s}{\sqrt{N_s+N_b}}\, .
\end{equation}
It  is then possible to obtain the luminosities needed to obtain an effect of a given statistical significance as :
\begin{equation}
L=\frac{S^2}{\sigma_s \epsilon_s} \left[1 + \frac{\sigma_b\epsilon_b}{\sigma_s \epsilon_s}\right]
\end{equation}
Therefore luminosity curves at 5- and 3-$\sigma$ level (i.e. fixing $S=3$ or $S=5$) can be straightforwardly given as a function of the mass $m_*$ of the exotic quark.
Fig.~\ref{fig:lumicurve} shows such 3- and 5 sigma luminosity curves which can also be used to get indications on the potential for discovery (or exclusion) at a given luminosity reached by the experiments at Run II of the LHC. 

\section{Conclusions \label{sec:disc_conc}}

We have presented the first study of  the production at the CERN LHC of new exotic quark states of charge $q=(5/3) e$ and $q=-(4/3)e$ which appear in composite models of quarks and leptons when considering higher isospin multiplets $I_W=1$ and $I_W=3/2$.
Such states have been discussed quite sometime back~\cite{Pancheri:1984sm} but their phenomenology has been, somewhat surprisingly, not addressed in detail. 
Only very recently~\cite{Biondini:2012ny,Biondini:2015aa,Leonardi:2014aa} some attention has been devoted to the phenomenology of exotic  doubly charged states appearing in the lepton sector of the extended weak isospin model of ref.~\cite{Pancheri:1984sm}.
Here we explore, to the best of our knowledge for the first time, the phenomenology of the hadron sector of the same model  with respect to the CERN LHC experiments, with a focus on the Run II at a center of mass energy of $\sqrt{s}=13$ TeV. This is the main motivation which started the present work which however acquires a relevant importance  in view of the fact that the model considered here has the potential of explaining, at least qualitatively, the  excess above the SM background reported very recently by the CMS collaboration in the analysis of the data of Run I at $\sqrt{s}=8$ TeV  in the $e\,\slashed{p}_T\, j\,j $ channel~\cite{CMS-PAS-EXO-12-041b,*Khachatryan:2016aa}.

\begin{table}[b!]
\begin{ruledtabular}
\begin{tabular}{ccc}
&3-$\sigma$ &5-$\sigma$\cr
\hline
L (fb$^{-1}$)  & $m_*$ (GeV)  & $m_*$ (GeV) \cr
\hline
30&  $2930 +70 -50$  & $2660+60-50$\cr
300&  $3540 +50 -30$ & $3280+60-50$\cr
3000&  4140 $\pm 60$& $3880+60-40$ 
\end{tabular}
\end{ruledtabular}
\caption{\label{tab:discoveryreach32} Discovery reach, in the case $I_W=3/2$, for $m_*$ at 3- and 5-$\sigma$ level within the statistical error at different values of the integrated luminosity  $L=(30,300,3000)$ fb$^{-1}$.}
\end{table}

This is particularly interesting in view of the fact that the recent studies~\cite{Leonardi:2016aa} of the lepton sector of extended weak-isospin composite multiplets suggest a possible explanation of the concomitant  CMS excess observed in the $eejj$ channel~\cite{Khachatryan:2014dka} in terms of an hypothetical composite Majorana neutrino.
In \cite{Leonardi:2016aa} it has been also suggested that the composite scenario could also be connected to the recent anomaly reported by the ATLAS collaboration  in a search of resonances in the di-boson channel~\cite{Aad:2015owa}. 

We have implemented the magnetic type Gauge interactions in the CalcHEP generator and performed a fast simulation of the detector reconstruction of both signal and relevant SM background ($Wjj$) based on the {\scshape{Delphes}} software~\cite{deFavereau:2013fsa}.

Finally we compute the luminosity curves as functions of $m_*$ for  3- and 5-$\sigma$ level statistical significance   including the statistical error. For different values of the integrated luminosity $L=(30,300,3000)$ fb$^{-1}$ we find for instance  that, for $I_W=3/2$ we can either observe or exclude (at a 3-$\sigma$ level) respectively masses up to $m_*\approx (2930,3540,4140)$ GeV. See table \ref{tab:discoveryreach32} \& \ref{tab:discoveryreach1} for details.

\begin{table}[b!]
\begin{ruledtabular}
\begin{tabular}{ccc}
&3-$\sigma$ &5-$\sigma$\cr
\hline
L (fb$^{-1}$)  & $m_*$ (GeV)  & $m_*$ (GeV) \cr
\hline
30&  $2230 +40 -60$  & $1980+40-50$\cr
300&  $2780 \pm 30$ & $2540+40-30$\cr
3000&  $3280 \pm 30$& $3060\pm 30$ 
\end{tabular}
\end{ruledtabular}
\caption{\label{tab:discoveryreach1} Discovery reach, in the case $I_W=1$, for $m_*$ at 3- and 5-$\sigma$ level within the statistical error at different values of the integrated luminosity  $L=(30,300,3000)$ fb$^{-1}$.}
\end{table}
 This is a quite interesting result which in our opinion warrants more detailed studies. For instance the two dimensional  parameter space ($\Lambda, m_*)$ could be fully explored. 
Also  the effect of expected contact interactions should be taken into account. This could improve the sensitivity of the signature to larger portions of the parameter space.

\begin{acknowledgments}
We acknowledge  useful discussions with the following members of the CMS Collaboration: L. Fan\`{o}, L. Alunni (University of Perugia) and  Francesco Romeo (IHEP, Beijing).
\end{acknowledgments}


\begin{thebibliography}{51}%
\makeatletter
\providecommand \@ifxundefined [1]{%
 \@ifx{#1\undefined}
}%
\providecommand \@ifnum [1]{%
 \ifnum #1\expandafter \@firstoftwo
 \else \expandafter \@secondoftwo
 \fi
}%
\providecommand \@ifx [1]{%
 \ifx #1\expandafter \@firstoftwo
 \else \expandafter \@secondoftwo
 \fi
}%
\providecommand \natexlab [1]{#1}%
\providecommand \enquote  [1]{``#1''}%
\providecommand \bibnamefont  [1]{#1}%
\providecommand \bibfnamefont [1]{#1}%
\providecommand \citenamefont [1]{#1}%
\providecommand \href@noop [0]{\@secondoftwo}%
\providecommand \href [0]{\begingroup \@sanitize@url \@href}%
\providecommand \@href[1]{\@@startlink{#1}\@@href}%
\providecommand \@@href[1]{\endgroup#1\@@endlink}%
\providecommand \@sanitize@url [0]{\catcode `\\12\catcode `\$12\catcode
  `\&12\catcode `\#12\catcode `\^12\catcode `\_12\catcode `\%12\relax}%
\providecommand \@@startlink[1]{}%
\providecommand \@@endlink[0]{}%
\providecommand \url  [0]{\begingroup\@sanitize@url \@url }%
\providecommand \@url [1]{\endgroup\@href {#1}{\urlprefix }}%
\providecommand \urlprefix  [0]{URL }%
\providecommand \Eprint [0]{\href }%
\providecommand \doibase [0]{http://dx.doi.org/}%
\providecommand \selectlanguage [0]{\@gobble}%
\providecommand \bibinfo  [0]{\@secondoftwo}%
\providecommand \bibfield  [0]{\@secondoftwo}%
\providecommand \translation [1]{[#1]}%
\providecommand \BibitemOpen [0]{}%
\providecommand \bibitemStop [0]{}%
\providecommand \bibitemNoStop [0]{.\EOS\space}%
\providecommand \EOS [0]{\spacefactor3000\relax}%
\providecommand \BibitemShut  [1]{\csname bibitem#1\endcsname}%
\let\auto@bib@innerbib\@empty
\bibitem [{\citenamefont {{Dirac}}(1963)}]{Dirac:1963aa}%
  \BibitemOpen
  \bibfield  {author} {\bibinfo {author} {\bibfnamefont {P.~A.~M.}\
  \bibnamefont {{Dirac}}},\ }\href {\doibase 10.1038/scientificamerican0563-45}
  {\bibfield  {journal} {\bibinfo  {journal} {Scientific American}\ }\textbf
  {\bibinfo {volume} {208}},\ \bibinfo {pages} {45} (\bibinfo {year}
  {1963})}\BibitemShut {NoStop}%
\bibitem [{\citenamefont {Terazawa}\ \emph {et~al.}(1977)\citenamefont
  {Terazawa}, \citenamefont {Akama},\ and\ \citenamefont
  {Chikashige}}]{Terazawa:1976xx}%
  \BibitemOpen
  \bibfield  {author} {\bibinfo {author} {\bibfnamefont {H.}~\bibnamefont
  {Terazawa}}, \bibinfo {author} {\bibfnamefont {K.}~\bibnamefont {Akama}}, \
  and\ \bibinfo {author} {\bibfnamefont {Y.}~\bibnamefont {Chikashige}},\
  }\href {\doibase 10.1103/PhysRevD.15.480} {\bibfield  {journal} {\bibinfo
  {journal} {Phys.Rev.}\ }\textbf {\bibinfo {volume} {D15}},\ \bibinfo {pages}
  {480} (\bibinfo {year} {1977})}\BibitemShut {NoStop}%
\bibitem [{\citenamefont {Terazawa}(1980)}]{Terazawa:1979pj}%
  \BibitemOpen
  \bibfield  {author} {\bibinfo {author} {\bibfnamefont {H.}~\bibnamefont
  {Terazawa}},\ }\href {\doibase 10.1103/PhysRevD.22.184} {\bibfield  {journal}
  {\bibinfo  {journal} {Phys.Rev.}\ }\textbf {\bibinfo {volume} {D22}},\
  \bibinfo {pages} {184} (\bibinfo {year} {1980})}\BibitemShut {NoStop}%
\bibitem [{\citenamefont {Eichten}\ and\ \citenamefont
  {Lane}(1980)}]{Eichten:1980aa}%
  \BibitemOpen
  \bibfield  {author} {\bibinfo {author} {\bibfnamefont {E.}~\bibnamefont
  {Eichten}}\ and\ \bibinfo {author} {\bibfnamefont {K.}~\bibnamefont {Lane}},\
  }\href {\doibase http://dx.doi.org/10.1016/0370-2693(80)90065-9} {\bibfield
  {journal} {\bibinfo  {journal} {Physics Letters B}\ }\textbf {\bibinfo
  {volume} {90}},\ \bibinfo {pages} {125 } (\bibinfo {year}
  {1980})}\BibitemShut {NoStop}%
\bibitem [{\citenamefont {Eichten}\ \emph {et~al.}(1983)\citenamefont
  {Eichten}, \citenamefont {Lane},\ and\ \citenamefont
  {Peskin}}]{Eichten:1983hw}%
  \BibitemOpen
  \bibfield  {author} {\bibinfo {author} {\bibfnamefont {E.}~\bibnamefont
  {Eichten}}, \bibinfo {author} {\bibfnamefont {K.~D.}\ \bibnamefont {Lane}}, \
  and\ \bibinfo {author} {\bibfnamefont {M.~E.}\ \bibnamefont {Peskin}},\
  }\href {\doibase 10.1103/PhysRevLett.50.811} {\bibfield  {journal} {\bibinfo
  {journal} {Phys. Rev. Lett.}\ }\textbf {\bibinfo {volume} {50}},\ \bibinfo
  {pages} {811} (\bibinfo {year} {1983})}\BibitemShut {NoStop}%
\bibitem [{\citenamefont {Terazawa}(1984)}]{Terazawa:1984bd}%
  \BibitemOpen
  \bibfield  {author} {\bibinfo {author} {\bibfnamefont {H.}~\bibnamefont
  {Terazawa}},\ }in\ \href@noop {} {\emph {\bibinfo {booktitle} {{Europhysics
  Topical Conference: Flavor Mixing in Weak Interactions Erice, Italy, March
  4-12, 1984}}}}\ (\bibinfo {year} {1984})\BibitemShut {NoStop}%
\bibitem [{\citenamefont {Cabibbo}\ \emph {et~al.}(1984)\citenamefont
  {Cabibbo}, \citenamefont {Maiani},\ and\ \citenamefont
  {Srivastava}}]{Cabibbo:1984aa}%
  \BibitemOpen
  \bibfield  {author} {\bibinfo {author} {\bibfnamefont {N.}~\bibnamefont
  {Cabibbo}}, \bibinfo {author} {\bibfnamefont {L.}~\bibnamefont {Maiani}}, \
  and\ \bibinfo {author} {\bibfnamefont {Y.}~\bibnamefont {Srivastava}},\
  }\href {\doibase 10.1016/0370-2693(84)91850-1} {\bibfield  {journal}
  {\bibinfo  {journal} {Phys. Lett.}\ }\textbf {\bibinfo {volume} {B139}},\
  \bibinfo {pages} {459} (\bibinfo {year} {1984})}\BibitemShut {NoStop}%
\bibitem [{\citenamefont {Baur}\ \emph {et~al.}(1990)\citenamefont {Baur},
  \citenamefont {Spira},\ and\ \citenamefont {Zerwas}}]{Baur:1990aa}%
  \BibitemOpen
  \bibfield  {author} {\bibinfo {author} {\bibfnamefont {U.}~\bibnamefont
  {Baur}}, \bibinfo {author} {\bibfnamefont {M.}~\bibnamefont {Spira}}, \ and\
  \bibinfo {author} {\bibfnamefont {P.~M.}\ \bibnamefont {Zerwas}},\ }\href
  {\doibase 10.1103/PhysRevD.42.815} {\bibfield  {journal} {\bibinfo  {journal}
  {Phys. Rev. D}\ }\textbf {\bibinfo {volume} {42}},\ \bibinfo {pages} {815}
  (\bibinfo {year} {1990})}\BibitemShut {NoStop}%
\bibitem [{\citenamefont {Baur}\ \emph {et~al.}(1987)\citenamefont {Baur},
  \citenamefont {Hinchliffe},\ and\ \citenamefont {Zeppenfeld}}]{Baur:1987ga}%
  \BibitemOpen
  \bibfield  {author} {\bibinfo {author} {\bibfnamefont {U.}~\bibnamefont
  {Baur}}, \bibinfo {author} {\bibfnamefont {I.}~\bibnamefont {Hinchliffe}}, \
  and\ \bibinfo {author} {\bibfnamefont {D.}~\bibnamefont {Zeppenfeld}},\
  }\href {\doibase 10.1142/S0217751X87000661} {\bibfield  {journal} {\bibinfo
  {journal} {Int. J. Mod. Phys.}\ }\textbf {\bibinfo {volume} {A2}},\ \bibinfo
  {pages} {1285} (\bibinfo {year} {1987})}\BibitemShut {NoStop}%
\bibitem [{\citenamefont {Aad}\ \emph {et~al.}(2010)\citenamefont {Aad} \emph
  {et~al.}}]{PhysRevLett.105.161801}%
  \BibitemOpen
  \bibfield  {author} {\bibinfo {author} {\bibfnamefont {G.}~\bibnamefont
  {Aad}} \emph {et~al.} (\bibinfo {collaboration} {ATLAS Collaboration}),\
  }\href {\doibase 10.1103/PhysRevLett.105.161801} {\bibfield  {journal}
  {\bibinfo  {journal} {Phys. Rev. Lett.}\ }\textbf {\bibinfo {volume} {105}},\
  \bibinfo {pages} {161801} (\bibinfo {year} {2010})}\BibitemShut {NoStop}%
\bibitem [{\citenamefont {Aad}\ \emph {et~al.}(2016{\natexlab{a}})\citenamefont
  {Aad} \emph {et~al.}}]{Aad:2016aa}%
  \BibitemOpen
  \bibfield  {author} {\bibinfo {author} {\bibfnamefont {G.}~\bibnamefont
  {Aad}} \emph {et~al.} (\bibinfo {collaboration} {ATLAS}),\ }\href {\doibase
  10.1007/JHEP02(2016)110} {\bibfield  {journal} {\bibinfo  {journal} {JHEP}\
  }\textbf {\bibinfo {volume} {02}},\ \bibinfo {pages} {110} (\bibinfo {year}
  {2016}{\natexlab{a}})},\ \Eprint {http://arxiv.org/abs/1510.02664}
  {arXiv:1510.02664 [hep-ex]} \BibitemShut {NoStop}%
\bibitem [{\citenamefont {Pancheri}\ and\ \citenamefont
  {Srivastava}(1984)}]{Pancheri:1984sm}%
  \BibitemOpen
  \bibfield  {author} {\bibinfo {author} {\bibfnamefont {G.}~\bibnamefont
  {Pancheri}}\ and\ \bibinfo {author} {\bibfnamefont {Y.~N.}\ \bibnamefont
  {Srivastava}},\ }\href {\doibase 10.1016/0370-2693(84)90649-X} {\bibfield
  {journal} {\bibinfo  {journal} {Phys. Lett.}\ }\textbf {\bibinfo {volume}
  {B146}},\ \bibinfo {pages} {87} (\bibinfo {year} {1984})}\BibitemShut
  {NoStop}%
\bibitem [{\citenamefont {Matsedonskyi}\ \emph {et~al.}(2016)\citenamefont
  {Matsedonskyi}, \citenamefont {Panico},\ and\ \citenamefont
  {Wulzer}}]{Matsedonskyi:2016aa}%
  \BibitemOpen
  \bibfield  {author} {\bibinfo {author} {\bibfnamefont {O.}~\bibnamefont
  {Matsedonskyi}}, \bibinfo {author} {\bibfnamefont {G.}~\bibnamefont
  {Panico}}, \ and\ \bibinfo {author} {\bibfnamefont {A.}~\bibnamefont
  {Wulzer}},\ }\href {\doibase 10.1007/JHEP04(2016)003} {\bibfield  {journal}
  {\bibinfo  {journal} {JHEP}\ }\textbf {\bibinfo {volume} {04}},\ \bibinfo
  {pages} {003} (\bibinfo {year} {2016})},\ \Eprint
  {http://arxiv.org/abs/1512.04356} {arXiv:1512.04356 [hep-ph]} \BibitemShut
  {NoStop}%
\bibitem [{\citenamefont {Aguila}\ and\ \citenamefont
  {Bowick}(1983)}]{Aguila:1983aa}%
  \BibitemOpen
  \bibfield  {author} {\bibinfo {author} {\bibfnamefont {F.~D.}\ \bibnamefont
  {Aguila}}\ and\ \bibinfo {author} {\bibfnamefont {M.}~\bibnamefont
  {Bowick}},\ }\href {\doibase http://dx.doi.org/10.1016/0550-3213(83)90316-4}
  {\bibfield  {journal} {\bibinfo  {journal} {Nuclear Physics B}\ }\textbf
  {\bibinfo {volume} {224}},\ \bibinfo {pages} {107 } (\bibinfo {year}
  {1983})}\BibitemShut {NoStop}%
\bibitem [{\citenamefont {Aguilar-Saavedra}(2009)}]{Aguilar-Saavedra:2009aa}%
  \BibitemOpen
  \bibfield  {author} {\bibinfo {author} {\bibfnamefont {J.~A.}\ \bibnamefont
  {Aguilar-Saavedra}},\ }\href
  {http://stacks.iop.org/1126-6708/2009/i=11/a=030} {\bibfield  {journal}
  {\bibinfo  {journal} {Journal of High Energy Physics}\ }\textbf {\bibinfo
  {volume} {2009}},\ \bibinfo {pages} {030} (\bibinfo {year}
  {2009})}\BibitemShut {NoStop}%
\bibitem [{\citenamefont {Aguilar-Saavedra}(2013)}]{Aguilar-Saavedra:2013aa}%
  \BibitemOpen
  \bibfield  {author} {\bibinfo {author} {\bibfnamefont {J.~A.}\ \bibnamefont
  {Aguilar-Saavedra}},\ }\href {\doibase 10.1051/epjconf/20136016012}
  {\bibfield  {journal} {\bibinfo  {journal} {EPJ Web of Conferences}\ }\textbf
  {\bibinfo {volume} {60}},\ \bibinfo {pages} {16012} (\bibinfo {year}
  {2013})}\BibitemShut {NoStop}%
\bibitem [{\citenamefont {Aguilar-Saavedra}\ \emph {et~al.}(2013)\citenamefont
  {Aguilar-Saavedra}, \citenamefont {Benbrik}, \citenamefont {Heinemeyer},\
  and\ \citenamefont {P\'erez-Victoria}}]{Aguilar-Saavedra:2013ab}%
  \BibitemOpen
  \bibfield  {author} {\bibinfo {author} {\bibfnamefont {J.~A.}\ \bibnamefont
  {Aguilar-Saavedra}}, \bibinfo {author} {\bibfnamefont {R.}~\bibnamefont
  {Benbrik}}, \bibinfo {author} {\bibfnamefont {S.}~\bibnamefont {Heinemeyer}},
  \ and\ \bibinfo {author} {\bibfnamefont {M.}~\bibnamefont
  {P\'erez-Victoria}},\ }\href {\doibase 10.1103/PhysRevD.88.094010} {\bibfield
   {journal} {\bibinfo  {journal} {Phys. Rev. D}\ }\textbf {\bibinfo {volume}
  {88}},\ \bibinfo {pages} {094010} (\bibinfo {year} {2013})}\BibitemShut
  {NoStop}%
\bibitem [{\citenamefont {Barducci}\ \emph {et~al.}(2015)\citenamefont
  {Barducci}, \citenamefont {Belyaev}, \citenamefont {Buchkremer},
  \citenamefont {Marrouche}, \citenamefont {Moretti},\ and\ \citenamefont
  {Panizzi}}]{Barducci:2015aa}%
  \BibitemOpen
  \bibfield  {author} {\bibinfo {author} {\bibfnamefont {D.}~\bibnamefont
  {Barducci}}, \bibinfo {author} {\bibfnamefont {A.}~\bibnamefont {Belyaev}},
  \bibinfo {author} {\bibfnamefont {M.}~\bibnamefont {Buchkremer}}, \bibinfo
  {author} {\bibfnamefont {J.}~\bibnamefont {Marrouche}}, \bibinfo {author}
  {\bibfnamefont {S.}~\bibnamefont {Moretti}}, \ and\ \bibinfo {author}
  {\bibfnamefont {L.}~\bibnamefont {Panizzi}},\ }\href {\doibase
  http://dx.doi.org/10.1016/j.cpc.2015.08.016} {\bibfield  {journal} {\bibinfo
  {journal} {Computer Physics Communications}\ }\textbf {\bibinfo {volume}
  {197}},\ \bibinfo {pages} {263 } (\bibinfo {year} {2015})}\BibitemShut
  {NoStop}%
\bibitem [{\citenamefont {Barducci}\ \emph {et~al.}(2014)\citenamefont
  {Barducci}, \citenamefont {Belyaev}, \citenamefont {Buchkremer},
  \citenamefont {Cacciapaglia}, \citenamefont {Deandrea}, \citenamefont
  {De~Curtis}, \citenamefont {Marrouche}, \citenamefont {Moretti},\ and\
  \citenamefont {Panizzi}}]{Barducci:2014aa}%
  \BibitemOpen
  \bibfield  {author} {\bibinfo {author} {\bibfnamefont {D.}~\bibnamefont
  {Barducci}}, \bibinfo {author} {\bibfnamefont {A.}~\bibnamefont {Belyaev}},
  \bibinfo {author} {\bibfnamefont {M.}~\bibnamefont {Buchkremer}}, \bibinfo
  {author} {\bibfnamefont {G.}~\bibnamefont {Cacciapaglia}}, \bibinfo {author}
  {\bibfnamefont {A.}~\bibnamefont {Deandrea}}, \bibinfo {author}
  {\bibfnamefont {S.}~\bibnamefont {De~Curtis}}, \bibinfo {author}
  {\bibfnamefont {J.}~\bibnamefont {Marrouche}}, \bibinfo {author}
  {\bibfnamefont {S.}~\bibnamefont {Moretti}}, \ and\ \bibinfo {author}
  {\bibfnamefont {L.}~\bibnamefont {Panizzi}},\ }\href {\doibase
  10.1007/JHEP12(2014)080} {\bibfield  {journal} {\bibinfo  {journal} {JHEP}\
  }\textbf {\bibinfo {volume} {12}},\ \bibinfo {pages} {080} (\bibinfo {year}
  {2014})},\ \Eprint {http://arxiv.org/abs/1405.0737} {arXiv:1405.0737
  [hep-ph]} \BibitemShut {NoStop}%
\bibitem [{\citenamefont {Cacciapaglia}\ \emph {et~al.}(2015)\citenamefont
  {Cacciapaglia}, \citenamefont {Deandrea}, \citenamefont {Gaur}, \citenamefont
  {Harada}, \citenamefont {Okada},\ and\ \citenamefont
  {Panizzi}}]{Cacciapaglia:2015aa}%
  \BibitemOpen
  \bibfield  {author} {\bibinfo {author} {\bibfnamefont {G.}~\bibnamefont
  {Cacciapaglia}}, \bibinfo {author} {\bibfnamefont {A.}~\bibnamefont
  {Deandrea}}, \bibinfo {author} {\bibfnamefont {N.}~\bibnamefont {Gaur}},
  \bibinfo {author} {\bibfnamefont {D.}~\bibnamefont {Harada}}, \bibinfo
  {author} {\bibfnamefont {Y.}~\bibnamefont {Okada}}, \ and\ \bibinfo {author}
  {\bibfnamefont {L.}~\bibnamefont {Panizzi}},\ }\href {\doibase
  10.1007/JHEP09(2015)012} {\bibfield  {journal} {\bibinfo  {journal} {JHEP}\
  }\textbf {\bibinfo {volume} {09}},\ \bibinfo {pages} {012} (\bibinfo {year}
  {2015})},\ \Eprint {http://arxiv.org/abs/1502.00370} {arXiv:1502.00370
  [hep-ph]} \BibitemShut {NoStop}%
\bibitem [{\citenamefont {Contino}\ and\ \citenamefont
  {Servant}(2008)}]{Contino:2008hi}%
  \BibitemOpen
  \bibfield  {author} {\bibinfo {author} {\bibfnamefont {R.}~\bibnamefont
  {Contino}}\ and\ \bibinfo {author} {\bibfnamefont {G.}~\bibnamefont
  {Servant}},\ }\href {\doibase 10.1088/1126-6708/2008/06/026} {\bibfield
  {journal} {\bibinfo  {journal} {JHEP}\ }\textbf {\bibinfo {volume} {06}},\
  \bibinfo {pages} {026} (\bibinfo {year} {2008})},\ \Eprint
  {http://arxiv.org/abs/0801.1679} {arXiv:0801.1679 [hep-ph]} \BibitemShut
  {NoStop}%
\bibitem [{\citenamefont {Atre}\ \emph {et~al.}(2011)\citenamefont {Atre},
  \citenamefont {Azuelos}, \citenamefont {Carena}, \citenamefont {Han},
  \citenamefont {Ozcan}, \citenamefont {Santiago},\ and\ \citenamefont
  {Unel}}]{Atre:2011ae}%
  \BibitemOpen
  \bibfield  {author} {\bibinfo {author} {\bibfnamefont {A.}~\bibnamefont
  {Atre}}, \bibinfo {author} {\bibfnamefont {G.}~\bibnamefont {Azuelos}},
  \bibinfo {author} {\bibfnamefont {M.}~\bibnamefont {Carena}}, \bibinfo
  {author} {\bibfnamefont {T.}~\bibnamefont {Han}}, \bibinfo {author}
  {\bibfnamefont {E.}~\bibnamefont {Ozcan}}, \bibinfo {author} {\bibfnamefont
  {J.}~\bibnamefont {Santiago}}, \ and\ \bibinfo {author} {\bibfnamefont
  {G.}~\bibnamefont {Unel}},\ }\href {\doibase 10.1007/JHEP08(2011)080}
  {\bibfield  {journal} {\bibinfo  {journal} {JHEP}\ }\textbf {\bibinfo
  {volume} {08}},\ \bibinfo {pages} {080} (\bibinfo {year} {2011})},\ \Eprint
  {http://arxiv.org/abs/1102.1987} {arXiv:1102.1987 [hep-ph]} \BibitemShut
  {NoStop}%
\bibitem [{\citenamefont {Chatrchyan}\ \emph {et~al.}(2014)\citenamefont
  {Chatrchyan} \emph {et~al.}}]{Chatrchyan:2014aa}%
  \BibitemOpen
  \bibfield  {author} {\bibinfo {author} {\bibfnamefont {S.}~\bibnamefont
  {Chatrchyan}} \emph {et~al.} (\bibinfo {collaboration} {CMS Collaboration}),\
  }\href {\doibase 10.1103/PhysRevLett.112.171801} {\bibfield  {journal}
  {\bibinfo  {journal} {Phys. Rev. Lett.}\ }\textbf {\bibinfo {volume} {112}},\
  \bibinfo {pages} {171801} (\bibinfo {year} {2014})}\BibitemShut {NoStop}%
\bibitem [{\citenamefont {Collaboration)}(2015)}]{CMS:2015alb}%
  \BibitemOpen
  \bibfield  {author} {\bibinfo {author} {\bibfnamefont {C.}~\bibnamefont
  {Collaboration)}} (\bibinfo {collaboration} {CMS Collaboration}),\
  }\href@noop {} {\ \textbf {\bibinfo {volume} {CMS-PAS-B2G-15-006}} (\bibinfo
  {year} {2015})}\BibitemShut {NoStop}%
\bibitem [{\citenamefont {Khachatryan}\ \emph {et~al.}(2010)\citenamefont
  {Khachatryan}, \citenamefont {Sirunyan}, \citenamefont {Tumasyan} \emph
  {et~al.}}]{PhysRevLett.105.211801}%
  \BibitemOpen
  \bibfield  {author} {\bibinfo {author} {\bibfnamefont {V.}~\bibnamefont
  {Khachatryan}}, \bibinfo {author} {\bibfnamefont {A.~M.}\ \bibnamefont
  {Sirunyan}}, \bibinfo {author} {\bibfnamefont {A.}~\bibnamefont {Tumasyan}},
  \emph {et~al.} (\bibinfo {collaboration} {CMS Collaboration}),\ }\href
  {\doibase 10.1103/PhysRevLett.105.211801} {\bibfield  {journal} {\bibinfo
  {journal} {Phys. Rev. Lett.}\ }\textbf {\bibinfo {volume} {105}},\ \bibinfo
  {pages} {211801} (\bibinfo {year} {2010})}\BibitemShut {NoStop}%
\bibitem [{\citenamefont {{CMS Collaboration}}(2016)}]{CMS-PAS-EXO-16-032}%
  \BibitemOpen
  \bibfield  {author} {\bibinfo {author} {\bibnamefont {{CMS Collaboration}}},\
  }\href {https://cds.cern.ch/record/2205150} {\emph {\bibinfo {title}
  {{Searches for narrow resonances decaying to dijets in proton-proton
  collisions at 13 TeV using 12.9 inverse femtobarns.}}}},\ \bibinfo {type}
  {CMS Physics Analysis Summary}\ (\bibinfo {year} {2016})\BibitemShut
  {NoStop}%
\bibitem [{\citenamefont {Khachatryan}\ \emph
  {et~al.}(2016{\natexlab{a}})\citenamefont {Khachatryan} \emph
  {et~al.}}]{Khachatryan:2016ab}%
  \BibitemOpen
  \bibfield  {author} {\bibinfo {author} {\bibfnamefont {V.}~\bibnamefont
  {Khachatryan}} \emph {et~al.} (\bibinfo {collaboration} {CMS
  Collaboration}),\ }\href {\doibase 10.1103/PhysRevLett.116.071801} {\bibfield
   {journal} {\bibinfo  {journal} {Phys. Rev. Lett.}\ }\textbf {\bibinfo
  {volume} {116}},\ \bibinfo {pages} {071801} (\bibinfo {year}
  {2016}{\natexlab{a}})}\BibitemShut {NoStop}%
\bibitem [{\citenamefont {Aad}\ \emph {et~al.}(2016{\natexlab{b}})\citenamefont
  {Aad} \emph {et~al.}}]{Aad:2016ab}%
  \BibitemOpen
  \bibfield  {author} {\bibinfo {author} {\bibfnamefont {G.}~\bibnamefont
  {Aad}} \emph {et~al.},\ }\href {\doibase
  http://dx.doi.org/10.1016/j.physletb.2016.01.032} {\bibfield  {journal}
  {\bibinfo  {journal} {Physics Letters B}\ }\textbf {\bibinfo {volume}
  {754}},\ \bibinfo {pages} {302 } (\bibinfo {year}
  {2016}{\natexlab{b}})}\BibitemShut {NoStop}%
\bibitem [{\citenamefont {Bhattacharya}\ \emph {et~al.}(2007)\citenamefont
  {Bhattacharya}, \citenamefont {Chauhan}, \citenamefont {Choudhary},\ and\
  \citenamefont {Choudhury}}]{Bhattacharya:2007aa}%
  \BibitemOpen
  \bibfield  {author} {\bibinfo {author} {\bibfnamefont {S.}~\bibnamefont
  {Bhattacharya}}, \bibinfo {author} {\bibfnamefont {S.~S.}\ \bibnamefont
  {Chauhan}}, \bibinfo {author} {\bibfnamefont {B.~C.}\ \bibnamefont
  {Choudhary}}, \ and\ \bibinfo {author} {\bibfnamefont {D.}~\bibnamefont
  {Choudhury}},\ }\href {\doibase 10.1103/PhysRevD.76.115017} {\bibfield
  {journal} {\bibinfo  {journal} {Phys. Rev. D}\ }\textbf {\bibinfo {volume}
  {76}},\ \bibinfo {pages} {115017} (\bibinfo {year} {2007})}\BibitemShut
  {NoStop}%
\bibitem [{\citenamefont {Bhattacharya}\ \emph {et~al.}(2009)\citenamefont
  {Bhattacharya}, \citenamefont {Chauhan}, \citenamefont {Choudhary},\ and\
  \citenamefont {Choudhury}}]{Bhattacharya:2009aa}%
  \BibitemOpen
  \bibfield  {author} {\bibinfo {author} {\bibfnamefont {S.}~\bibnamefont
  {Bhattacharya}}, \bibinfo {author} {\bibfnamefont {S.~S.}\ \bibnamefont
  {Chauhan}}, \bibinfo {author} {\bibfnamefont {B.~C.}\ \bibnamefont
  {Choudhary}}, \ and\ \bibinfo {author} {\bibfnamefont {D.}~\bibnamefont
  {Choudhury}},\ }\href {\doibase 10.1103/PhysRevD.80.015014} {\bibfield
  {journal} {\bibinfo  {journal} {Phys. Rev. D}\ }\textbf {\bibinfo {volume}
  {80}},\ \bibinfo {pages} {015014} (\bibinfo {year} {2009})}\BibitemShut
  {NoStop}%
\bibitem [{\citenamefont {Golling}(2016)}]{Golling:2016aa}%
  \BibitemOpen
  \bibfield  {author} {\bibinfo {author} {\bibfnamefont {T.}~\bibnamefont
  {Golling}},\ }\href {\doibase http://dx.doi.org/10.1016/j.ppnp.2016.05.001}
  {\bibfield  {journal} {\bibinfo  {journal} {Progress in Particle and Nuclear
  Physics}\ }\textbf {\bibinfo {volume} {90}},\ \bibinfo {pages} {156 }
  (\bibinfo {year} {2016})}\BibitemShut {NoStop}%
\bibitem [{\citenamefont {Biondini}\ \emph {et~al.}(2012)\citenamefont
  {Biondini}, \citenamefont {Panella}, \citenamefont {Pancheri}, \citenamefont
  {Srivastava},\ and\ \citenamefont {Fan\`{o}}}]{Biondini:2012ny}%
  \BibitemOpen
  \bibfield  {author} {\bibinfo {author} {\bibfnamefont {S.}~\bibnamefont
  {Biondini}}, \bibinfo {author} {\bibfnamefont {O.}~\bibnamefont {Panella}},
  \bibinfo {author} {\bibfnamefont {G.}~\bibnamefont {Pancheri}}, \bibinfo
  {author} {\bibfnamefont {Y.}~\bibnamefont {Srivastava}}, \ and\ \bibinfo
  {author} {\bibfnamefont {L.}~\bibnamefont {Fan\`{o}}},\ }\href {\doibase
  10.1103/PhysRevD.85.095018} {\bibfield  {journal} {\bibinfo  {journal} {Phys.
  Rev. D}\ }\textbf {\bibinfo {volume} {85}},\ \bibinfo {pages} {095018}
  (\bibinfo {year} {2012})},\ \Eprint {http://arxiv.org/abs/1201.3764}
  {arXiv:1201.3764 [hep-ph]} \BibitemShut {NoStop}%
\bibitem [{\citenamefont {Leonardi}\ \emph {et~al.}(2014)\citenamefont
  {Leonardi}, \citenamefont {Panella},\ and\ \citenamefont
  {Fan\`o}}]{Leonardi:2014aa}%
  \BibitemOpen
  \bibfield  {author} {\bibinfo {author} {\bibfnamefont {R.}~\bibnamefont
  {Leonardi}}, \bibinfo {author} {\bibfnamefont {O.}~\bibnamefont {Panella}}, \
  and\ \bibinfo {author} {\bibfnamefont {L.}~\bibnamefont {Fan\`o}},\ }\href
  {\doibase 10.1103/PhysRevD.90.035001} {\bibfield  {journal} {\bibinfo
  {journal} {Phys. Rev. D}\ }\textbf {\bibinfo {volume} {90}},\ \bibinfo
  {pages} {035001} (\bibinfo {year} {2014})}\BibitemShut {NoStop}%
\bibitem [{\citenamefont {Biondini}\ and\ \citenamefont
  {Panella}(2015)}]{Biondini:2015aa}%
  \BibitemOpen
  \bibfield  {author} {\bibinfo {author} {\bibfnamefont {S.}~\bibnamefont
  {Biondini}}\ and\ \bibinfo {author} {\bibfnamefont {O.}~\bibnamefont
  {Panella}},\ }\href {\doibase 10.1103/PhysRevD.92.015023} {\bibfield
  {journal} {\bibinfo  {journal} {Phys. Rev. D}\ }\textbf {\bibinfo {volume}
  {92}},\ \bibinfo {pages} {015023} (\bibinfo {year} {2015})}\BibitemShut
  {NoStop}%
\bibitem [{\citenamefont {Leonardi}\ \emph {et~al.}(2016)\citenamefont
  {Leonardi}, \citenamefont {Alunni}, \citenamefont {Romeo}, \citenamefont
  {Fan{\`o}},\ and\ \citenamefont {Panella}}]{Leonardi:2016aa}%
  \BibitemOpen
  \bibfield  {author} {\bibinfo {author} {\bibfnamefont {R.}~\bibnamefont
  {Leonardi}}, \bibinfo {author} {\bibfnamefont {L.}~\bibnamefont {Alunni}},
  \bibinfo {author} {\bibfnamefont {F.}~\bibnamefont {Romeo}}, \bibinfo
  {author} {\bibfnamefont {L.}~\bibnamefont {Fan{\`o}}}, \ and\ \bibinfo
  {author} {\bibfnamefont {O.}~\bibnamefont {Panella}},\ }\href {\doibase
  10.1140/epjc/s10052-016-4396-y} {\bibfield  {journal} {\bibinfo  {journal}
  {The European Physical Journal C}\ }\textbf {\bibinfo {volume} {76}},\
  \bibinfo {pages} {593} (\bibinfo {year} {2016})}\BibitemShut {NoStop}%
\bibitem [{CMS(2016)}]{CMS-PAS-EXO-16-026}%
  \BibitemOpen
  \href {https://cds.cern.ch/record/2205277} {\emph {\bibinfo {title} {{Search
  for heavy composite Majorana neutrinos produced in association with a lepton
  and decaying into a same-flavour lepton plus two quarks at $\sqrt{s} =
  13~\mathrm{TeV}$ with the CMS detector}}}},\ \bibinfo {type} {Tech. Rep.}\
  \bibinfo {number} {CMS-PAS-EXO-16-026}\ (\bibinfo  {institution} {CERN},\
  \bibinfo {address} {Geneva},\ \bibinfo {year} {2016})\BibitemShut {NoStop}%
\bibitem [{\citenamefont {Khachatryan}\ \emph {et~al.}(2014)\citenamefont
  {Khachatryan} \emph {et~al.}}]{Khachatryan:2014dka}%
  \BibitemOpen
  \bibfield  {author} {\bibinfo {author} {\bibfnamefont {V.}~\bibnamefont
  {Khachatryan}} \emph {et~al.} (\bibinfo {collaboration} {CMS
  Collaboration}),\ }\href {\doibase 10.1140/epjc/s10052-014-3149-z} {\bibfield
   {journal} {\bibinfo  {journal} {Eur.Phys.J.}\ }\textbf {\bibinfo {volume}
  {C74}},\ \bibinfo {pages} {3149} (\bibinfo {year} {2014})},\ \Eprint
  {http://arxiv.org/abs/1407.3683} {arXiv:1407.3683 [hep-ex]} \BibitemShut
  {NoStop}%
\bibitem [{CMS(2014)}]{CMS-PAS-EXO-12-041b}%
  \BibitemOpen
  \href@noop {} {\emph {\bibinfo {title} {{Search for Pair-production of First
  Generation Scalar Leptoquarks in pp Collisions at $\sqrt{s} = 8$ TeV}}}},\
  \bibinfo {type} {Tech. Rep.}\ \bibinfo {number} {CMS-PAS-EXO-12-041}\
  (\bibinfo  {institution} {CERN},\ \bibinfo {address} {Geneva},\ \bibinfo
  {year} {2014})\BibitemShut {NoStop}%
\bibitem [{\citenamefont {Khachatryan}\ \emph
  {et~al.}(2016{\natexlab{b}})\citenamefont {Khachatryan} \emph
  {et~al.}}]{Khachatryan:2016aa}%
  \BibitemOpen
  \bibfield  {author} {\bibinfo {author} {\bibfnamefont {V.}~\bibnamefont
  {Khachatryan}} \emph {et~al.} (\bibinfo {collaboration} {CMS
  Collaboration}),\ }\href {\doibase 10.1103/PhysRevD.93.032004} {\bibfield
  {journal} {\bibinfo  {journal} {Phys. Rev. D}\ }\textbf {\bibinfo {volume}
  {93}},\ \bibinfo {pages} {032004} (\bibinfo {year}
  {2016}{\natexlab{b}})}\BibitemShut {NoStop}%
\bibitem [{\citenamefont {de~Favereau}\ \emph {et~al.}(2014)\citenamefont
  {de~Favereau} \emph {et~al.}}]{deFavereau:2013fsa}%
  \BibitemOpen
  \bibfield  {author} {\bibinfo {author} {\bibfnamefont {J.}~\bibnamefont
  {de~Favereau}} \emph {et~al.} (\bibinfo {collaboration} {DELPHES 3}),\ }\href
  {\doibase 10.1007/JHEP02(2014)057} {\bibfield  {journal} {\bibinfo  {journal}
  {JHEP}\ }\textbf {\bibinfo {volume} {1402}},\ \bibinfo {pages} {057}
  (\bibinfo {year} {2014})},\ \Eprint {http://arxiv.org/abs/1307.6346}
  {arXiv:1307.6346 [hep-ex]} \BibitemShut {NoStop}%
\bibitem [{\citenamefont {Aad}\ and\ \citenamefont
  {Zwalinski}(2013)}]{Aad:2013ab}%
  \BibitemOpen
  \bibfield  {author} {\bibinfo {author} {\bibfnamefont {G.}~\bibnamefont
  {Aad}}\ and\ \bibinfo {author} {\bibfnamefont {L.}~\bibnamefont
  {Zwalinski}},\ }\href {http://stacks.iop.org/1367-2630/15/i=9/a=093011}
  {\bibfield  {journal} {\bibinfo  {journal} {New Journal of Physics}\ }\textbf
  {\bibinfo {volume} {15}},\ \bibinfo {pages} {093011} (\bibinfo {year}
  {2013})}\BibitemShut {NoStop}%
\bibitem [{CMS(2015)}]{CMS:2015jga}%
  \BibitemOpen
  \href@noop {} {\emph {\bibinfo {title} {{Search for excited leptons in proton
  proton collisions at $\sqrt{s}$=8} TeV}}},\ \bibinfo {type} {Tech. Rep.}\
  (\bibinfo {year} {2015})\BibitemShut {NoStop}%
\bibitem [{\citenamefont {Pukhov}(2004)}]{Pukhov:2004ca}%
  \BibitemOpen
  \bibfield  {author} {\bibinfo {author} {\bibfnamefont {A.}~\bibnamefont
  {Pukhov}},\ }\href@noop {} {\  (\bibinfo {year} {2004})},\ \Eprint
  {http://arxiv.org/abs/hep-ph/0412191} {arXiv:hep-ph/0412191} \BibitemShut
  {NoStop}%
\bibitem [{\citenamefont {Christensen}\ and\ \citenamefont
  {Duhr}(2009)}]{Christensen:2008py}%
  \BibitemOpen
  \bibfield  {author} {\bibinfo {author} {\bibfnamefont {N.~D.}\ \bibnamefont
  {Christensen}}\ and\ \bibinfo {author} {\bibfnamefont {C.}~\bibnamefont
  {Duhr}},\ }\href {\doibase 10.1016/j.cpc.2009.02.018} {\bibfield  {journal}
  {\bibinfo  {journal} {Comput. Phys. Commun.}\ }\textbf {\bibinfo {volume}
  {180}},\ \bibinfo {pages} {1614} (\bibinfo {year} {2009})},\ \Eprint
  {http://arxiv.org/abs/0806.4194} {arXiv:0806.4194 [hep-ph]} \BibitemShut
  {NoStop}%
\bibitem [{\citenamefont {Belyaev}\ \emph {et~al.}(1999)\citenamefont
  {Belyaev}, \citenamefont {Boos},\ and\ \citenamefont
  {Dudko}}]{Belyaev:1998dn}%
  \BibitemOpen
  \bibfield  {author} {\bibinfo {author} {\bibfnamefont {A.~S.}\ \bibnamefont
  {Belyaev}}, \bibinfo {author} {\bibfnamefont {E.~E.}\ \bibnamefont {Boos}}, \
  and\ \bibinfo {author} {\bibfnamefont {L.~V.}\ \bibnamefont {Dudko}},\ }\href
  {\doibase 10.1103/PhysRevD.59.075001} {\bibfield  {journal} {\bibinfo
  {journal} {Phys. Rev.}\ }\textbf {\bibinfo {volume} {D59}},\ \bibinfo {pages}
  {075001} (\bibinfo {year} {1999})},\ \Eprint
  {http://arxiv.org/abs/hep-ph/9806332} {arXiv:hep-ph/9806332} \BibitemShut
  {NoStop}%
\bibitem [{\citenamefont {Belyaev}\ \emph {et~al.}(1995)\citenamefont
  {Belyaev}, \citenamefont {Boos}, \citenamefont {Dudko},\ and\ \citenamefont
  {Pukhov}}]{Belyaev:1995yu}%
  \BibitemOpen
  \bibfield  {author} {\bibinfo {author} {\bibfnamefont {A.}~\bibnamefont
  {Belyaev}}, \bibinfo {author} {\bibfnamefont {E.}~\bibnamefont {Boos}},
  \bibinfo {author} {\bibfnamefont {L.}~\bibnamefont {Dudko}}, \ and\ \bibinfo
  {author} {\bibfnamefont {A.}~\bibnamefont {Pukhov}},\ }\href@noop {} {\
  (\bibinfo {year} {1995})},\ \Eprint {http://arxiv.org/abs/hep-ph/9511306}
  {arXiv:hep-ph/9511306} \BibitemShut {NoStop}%
\bibitem [{\citenamefont {Eriksson}\ \emph {et~al.}(2008)\citenamefont
  {Eriksson}, \citenamefont {Hesselbach},\ and\ \citenamefont
  {Rathsman}}]{Eriksson:2006yt}%
  \BibitemOpen
  \bibfield  {author} {\bibinfo {author} {\bibfnamefont {D.}~\bibnamefont
  {Eriksson}}, \bibinfo {author} {\bibfnamefont {S.}~\bibnamefont
  {Hesselbach}}, \ and\ \bibinfo {author} {\bibfnamefont {J.}~\bibnamefont
  {Rathsman}},\ }\href {\doibase 10.1140/epjc/s10052-007-0453-x} {\bibfield
  {journal} {\bibinfo  {journal} {Eur. Phys. J.}\ }\textbf {\bibinfo {volume}
  {C53}},\ \bibinfo {pages} {267} (\bibinfo {year} {2008})},\ \Eprint
  {http://arxiv.org/abs/hep-ph/0612198} {arXiv:hep-ph/0612198} \BibitemShut
  {NoStop}%
\bibitem [{\citenamefont {Eriksson}\ \emph {et~al.}(2007)\citenamefont
  {Eriksson}, \citenamefont {Hesselbach},\ and\ \citenamefont
  {Rathsman}}]{Eriksson:2007re}%
  \BibitemOpen
  \bibfield  {author} {\bibinfo {author} {\bibfnamefont {D.}~\bibnamefont
  {Eriksson}}, \bibinfo {author} {\bibfnamefont {S.}~\bibnamefont
  {Hesselbach}}, \ and\ \bibinfo {author} {\bibfnamefont {J.}~\bibnamefont
  {Rathsman}},\ }\href@noop {} {\  (\bibinfo {year} {2007})},\ \Eprint
  {http://arxiv.org/abs/0710.3346} {arXiv:0710.3346 [hep-ph]} \BibitemShut
  {NoStop}%
\bibitem [{\citenamefont {Ball}\ \emph {et~al.}(2015)\citenamefont {Ball},
  \citenamefont {Bertone}, \citenamefont {Carrazza}, \citenamefont {Deans},
  \citenamefont {Del~Debbio}, \citenamefont {Forte}, \citenamefont {Guffanti},
  \citenamefont {Hartland}, \citenamefont {Latorre}, \citenamefont {Rojo},\
  and\ \citenamefont {Ubiali}}]{Ball:2015aa}%
  \BibitemOpen
  \bibfield  {author} {\bibinfo {author} {\bibfnamefont {R.~D.}\ \bibnamefont
  {Ball}}, \bibinfo {author} {\bibfnamefont {V.}~\bibnamefont {Bertone}},
  \bibinfo {author} {\bibfnamefont {S.}~\bibnamefont {Carrazza}}, \bibinfo
  {author} {\bibfnamefont {C.~S.}\ \bibnamefont {Deans}}, \bibinfo {author}
  {\bibfnamefont {L.}~\bibnamefont {Del~Debbio}}, \bibinfo {author}
  {\bibfnamefont {S.}~\bibnamefont {Forte}}, \bibinfo {author} {\bibfnamefont
  {A.}~\bibnamefont {Guffanti}}, \bibinfo {author} {\bibfnamefont {N.~P.}\
  \bibnamefont {Hartland}}, \bibinfo {author} {\bibfnamefont {J.~I.}\
  \bibnamefont {Latorre}}, \bibinfo {author} {\bibfnamefont {J.}~\bibnamefont
  {Rojo}}, \ and\ \bibinfo {author} {\bibfnamefont {M.}~\bibnamefont
  {Ubiali}},\ }\href {\doibase 10.1007/JHEP04(2015)040} {\bibfield  {journal}
  {\bibinfo  {journal} {Journal of High Energy Physics}\ }\textbf {\bibinfo
  {volume} {04}},\ \bibinfo {pages} {040} (\bibinfo {year} {2015})}\BibitemShut
  {NoStop}%
\bibitem [{\citenamefont {Richter}(2013)}]{Richter:2013aa}%
  \BibitemOpen
  \bibfield  {author} {\bibinfo {author} {\bibfnamefont {S.}~\bibnamefont
  {Richter}},\ }\emph {\bibinfo {title} {{Invariant mass reconstruction in a
  search for light charged Higgs bosons in $pp$ collisions at $\sqrt{s}$ = 7
  TeV}}},\ \href
  {https://inspirehep.net/record/1296475/files/549236788_MSc_thesis_Stefan_Richter_with_info_pages.pdf}
  {Master's thesis},\ \bibinfo  {school} {Helsinki U.} (\bibinfo {year}
  {2013})\BibitemShut {NoStop}%
\bibitem [{\citenamefont {Aad}\ \emph {et~al.}(2015)\citenamefont {Aad} \emph
  {et~al.}}]{Aad:2015owa}%
  \BibitemOpen
  \bibfield  {author} {\bibinfo {author} {\bibfnamefont {G.}~\bibnamefont
  {Aad}} \emph {et~al.} (\bibinfo {collaboration} {ATLAS}),\ }\href {\doibase
  10.1007/JHEP12(2015)055} {\bibfield  {journal} {\bibinfo  {journal} {JHEP}\
  }\textbf {\bibinfo {volume} {12}},\ \bibinfo {pages} {055} (\bibinfo {year}
  {2015})},\ \Eprint {http://arxiv.org/abs/1506.00962} {arXiv:1506.00962
  [hep-ex]} \BibitemShut {NoStop}%
\end{thebibliography}

%

\end{document}